\documentclass[useAMS,usenatbib]{mn2e}

\usepackage{amsmath}
\usepackage{amssymb}
\usepackage[usenames,dvipsnames]{xcolor}
\usepackage{graphicx}
\usepackage{graphics}
\def\apj{{ ApJ}}
\def\apjl{{ApJL}}
\def\apjs{{ ApJS}}
\def\apss{{ Ap\&SS}}
\def\aap{{ A\&A}}
\def\aj{{ AJ}}
\def\mnras{{ MNRAS}}
\def\araa{{ ARA\&A}}

\def\nat {{ Nature}}

\def\physrep{{Physics Reports}}

\def\ttheta{{\tilde{\theta}}}
\def\tphi{{\tilde{\phi}}}
\def\thetap{{\theta^\prime}}
\def\phip{{\phi^\prime}}
\def\nup{{\nu^\prime}}
\def\ggam{{\gamma\gamma}}
\newcommand{\myemail}{wenbinlu@astro.as.utexas.edu}

\title{Probing Massive Stars around Gamma-Ray Burst Progenitors}
\author[Lu, Kumar \& Smoot]
  {Wenbin Lu,$^1$\thanks{\myemail}
  Pawan Kumar,$^1$\thanks{pk@astro.as.utexas.edu}
  George F. Smoot$^2$\thanks{gfsmoot@lbl.gov}\\
  $^1$Department of Astronomy, University of Texas at Austin, Austin,
TX 78712, USA\\
  $^2$PCCP; APC, Universit\'{e} Paris Diderot,
  Universit\'{e} Sorbonne Paris Cit\'{e}, 75013 Paris, France\\
  $^2$BCCP; LBNL \& Physics Dept. University of California
  at Berkeley CA 94720, USA}
\date{\today}

\pagerange{\pageref{firstpage}--\pageref{lastpage}} \pubyear{2015}

\def\LaTeX{L\kern-.36em\raise.3ex\hbox{a}\kern-.15em
    T\kern-.1667em\lower.7ex\hbox{E}\kern-.125emX}

\begin{document}

\label{firstpage}

\maketitle

\begin{abstract}
Long Gamma-Ray Bursts (GRBs) are produced by ultra-relativistic jets
launched from core collapse of massive stars. 
Most massive stars form in binaries and/or in star clusters,
which means that there may be a significant external photon field (EPF) around
the GRB progenitor. We calculate the inverse-Compton scattering
of EPF by the hot electrons in the GRB jet. Three possible cases of EPF are
considered: the progenitor is {\it (I)} in a massive binary system,
{\it (II)} surrounded by a Wolf-Rayet-star wind, and  {\it (III)} in a dense star
cluster. Typical luminosities of $10^{46}-10^{50}$ $erg/s$ in the 
 $1$ - $100$ $GeV$ band are expected, depending 
on the stellar luminosity, binary separation  {\it (I)},
wind mass loss rate  {\it (II)}, stellar number density  {\it (III)},
etc. We calculate the lightcurve and spectrum in each case, taking
fully into account the equal-arrival time surfaces and possible pair-production
absorption with the prompt $\gamma$-rays. Observations can put constraints
on the existence of such EPFs (and hence on the nature of GRB
progenitors) and on the radius where the jet internal dissipation process
accelerates electrons.
\end{abstract}


\section{Introduction}
Long Gamma-Ray Bursts (GRBs) are produced by ultra-relativistic jets
launched after the collapse of massive stars
\citep[e.g.][]{1998Natur.395..670G, 
  2003Natur.423..847H, 2004RvMP...76.1143P, 2006ARA&A..44..507W}.
Massive stars tend to form in binaries and/or in dense star clusters
\citep{1998ApJ...493..180M,
  2007ARA&A..45..481Z, 2009AJ....137.3358M, 2010ARA&A..48..431P},
which produce a dense external photon field (EPF) around the GRB
progenitor. Soft photons will be inverse-Compton (IC) scattered by
electrons in the jet. The scattering will boost the energy of photons
by a factor of $\Gamma^2\gamma_e^2$, where $\Gamma\sim 300$ is the jet
bulk Lorentz factor (LF) and $\gamma_e$ is electrons' LF in the
comoving frame. Therefore, we expect EPF at $10\mbox{ }eV$ being
scattered to $10^6\gamma_e^2\ eV$.

In the GRB literature, this external inverse-Compton
(EIC) emission has been considered for a variety of sources of soft
photons and scattering electrons. For example, the soft photons may
come from a pulsar wind bubble created before the GRB
\citep{2003MNRAS.340..115G}, and from the funnel of the collapsing star created
during the jet breakout \citep{2000ApJ...529L..17L, 2000MNRAS.316L..45G,
2001ApJ...550..410M}. \citet{2004MNRAS.349L..38R} considered soft
photons from a companion
star being IC scattered by a magnetically driven relativistic wind
from a spinning-down millisecond
supra-magnetar. \citet{2008A&A...488L..55G, 2011MNRAS.418..583M}
considered the GRB external shock interacting with the photon field
from an O star in the surrounding dense cluster.

Different from previous studies, we consider
three cases where significant EPFs exist: the GRB progenitor {\it (I)} is in
a massive binary system, {\it (II)} has a strong wind, and {\it (III)}
is in a dense star cluster.

In {\it Case (I)}, soft photons come from a massive companion
star. Spectroscopic and direct imaging studies of Galactic massive
stars, although biased by selection effects and measurement
limitations, conclude that massive 
stars have binary fraction of at least $\sim50\%$ \citep[see][and
reference therein]{2009AJ....137.3358M}. Therefore, a companion of
comparable mass may exist near the GRB progenitor, depending on the
evolution status of the binary system \citep{2012ARA&A..50..107L}. For
binary separation $d = 10^{15}\mbox{ }cm$ and the companion star's
bolometric luminosity $L_b=10^{39}\mbox{ }erg/s$, the number of
scattered photons will be $10^{52} -10^{53}$ (isotropic equivalent).

In {\it Case (II)}, soft photons originally from
the progenitor star are scattered first by the electrons in the
stellar wind and again by the jet. A compact Wolf-Rayet (W-R) star
progenitor is favored 
by considering the propagation time of a GRB jet in the stellar envelope
\citep{2003MNRAS.345..575M}. W-R stars have high luminosities
of $\sim10^{39}\mbox{ }erg/s$ and are surrounded by strong stellar
wind of a typical mass loss rate $10^{-5}\mbox{ }M_{\odot}/yr$
\citep{2007ARA&A..45..177C}. The number of scattered photons is
$10^{50}-10^{51}$ (isotropic equivalent).

In {\it Case (III)}, all the stars in the cluster create a nearly
isotropic and uniform EPF around the GRB jet. Massive stars form in
clustered environments \citep{2003ARA&A..41...57L}. In our Galaxy,
$\sim85\%$ of O stars are either observed in young clusters or
directly identified as runaways \citep[see][and reference
therein]{2010ARA&A..48..431P}. Within the central $\sim0.2\mbox{ }pc$
of Arches Cluster, the densest known 
young massive cluster in the Milky Way, the number density of massive
($30<M/M_\odot<120$) stars is $\sim5000 
\mbox{ }pc^{-3}$\citep{2009A&A...501..563E}. These stars have 
luminosities close to or higher than $10^{39}\mbox{ }erg/s$.
Long GRBs are associated with actively star-forming 
galaxies \citep{2006Natur.441..463F, 2006ARA&A..44..507W}, where young
massive clusters are particularly abundant
\citep{2010ARA&A..48..431P}. Therefore, we expect a good fraction,
depending on the clusters' evolution, of long GRBs to happen in such
star clusters. In this {\it Case}, we consider EIC emission from both
internal dissipation and external shocks (both reverse shock and forward
shock). Note that our model is different from
\citet{2008A&A...488L..55G}, who consider the case when the external
forward shock happens to sweep across one star in the
cluster. \citet{2011MNRAS.418..583M} also calculate the EIC scattering
of the isotropic diffuse EPF in the cluster, but only the external forward
shock is considered. As shown in Section 3.1.3, the EIC luminosity
is dominated by the internal dissipation and reverse shock.

In this paper, we show that all three cases may have
observable consequences, which can be used to probe massive stars
around GRB progenitors and the nature of GRB progenitors.

A few more notes. (1) By ``internal dissipation'' (ID) we mean the
uncertain process \citep[e.g. internal shock or magnetic
reconnection, see][for a recent
review]{2015PhR...561....1K} that dissipates the free energy of the
jet and produces the prompt $\gamma$-rays.  As long as electrons are
accelerated to an ultra-relativistic powerlaw, our model is independent of
the details of the jet dissipation and particle acceleration process.
(2) The opening angle of the jet is assumed to be
larger than the causally connected cone,
i.e. $\theta_{j,max}>1/\Gamma$, so we use 
 the isotropic equivalent energy, luminosity, photon number, etc
 throughout the paper. (3) The jet is assumed to be hadron dominated
 and made of pure 
protons and electrons\footnote{Hadrons carry the momentum of the jet
  but are neglected in the IC scattering process (since there is an
  equal amount of electrons). If the jet is
  magnetically dominated, 
    we expect the EIC emission from the internal dissipation to be much
    weaker, due to a much smaller number of electrons. For a
    multi-element jet, e.g. Hydrogen 
    and Helium, our model can be easily modified by putting a
    $ (1+X(H))/2$ factor before $E_j$ (the total kinetic energy of the
    jet), where $X(H)$ is the mass fraction of Hydrogen.}.
(4) The convention $X_n = X/10^n$ in CGS units is adopted.

This paper is organized as follows. In Section 2, we give a general
treatment of the EIC emission (spectrum and lightcurve) when a jet
interacts with an arbitrary EPF. In Section 3, we first give simple
order-of-magnitude estimates of the 
total EIC energy in the three cases, and then put the
specific EPFs into the general procedures and calculate the precise
lightcurves and spectra. In Section 4, we discuss some
potential issues, e.g. absorption of high energy $\gamma$-rays by the
cosmic background light or by the local EPF itself. Short conclusion
is given in Section 5.

\section{Modeling the EIC Emission}\label{section:theory}
In this section, we provide a general calculation for the lightcurve
and spectrum expected for an observer on the jet axis, when a
relativistic jet makes its way through an arbitrary external photon
field (EPF).

EIC emission has been studied extensively in the literature, for
instance in active galactic nuclei
\citep[e.g.][]{1993ApJ...416..458D} where a jet meets an EPF. We refer
the readers to \citet{1970RvMP...42..237B, 1981Ap&SS..79..321A} for a
general treatment on IC radiation. Also, \citet{2008MNRAS.384.1483F}
give a review of IC radiation in GRB afterglows,
including EIC and synchrotron-self Compton (SSC) radiation. Our
treatment is different only in that, at an early stage, equal-arrival time
surfaces are taken into consideration. Experienced readers could skip
Section 2 and go to Section 3 for the results.

In terms of notations, (1) we use un-primed quantities in the lab frame
(rest frame of the progenitor star) and primed ($^\prime$) ones in the
jet comoving frame, and quantities with subscript ``$j$'',
e.g. $(\theta_j,\phi_j)$, are related to different regions 
in the jet. Note that, by ``comoving frame'' we mean ``comoving'' with
the bulk motion of the jet, rather than the host galaxy's
redshift. (2) We denote the unscattered and scattered photons'
frequencies as $\nu_0$ ($\nu_0^\prime$) and $\nu$ ($\nu^\prime$) in
the lab (comoving) frame.

\begin{figure}
  \centering
  \includegraphics[width=0.35\textwidth,
  height=0.23\textheight]{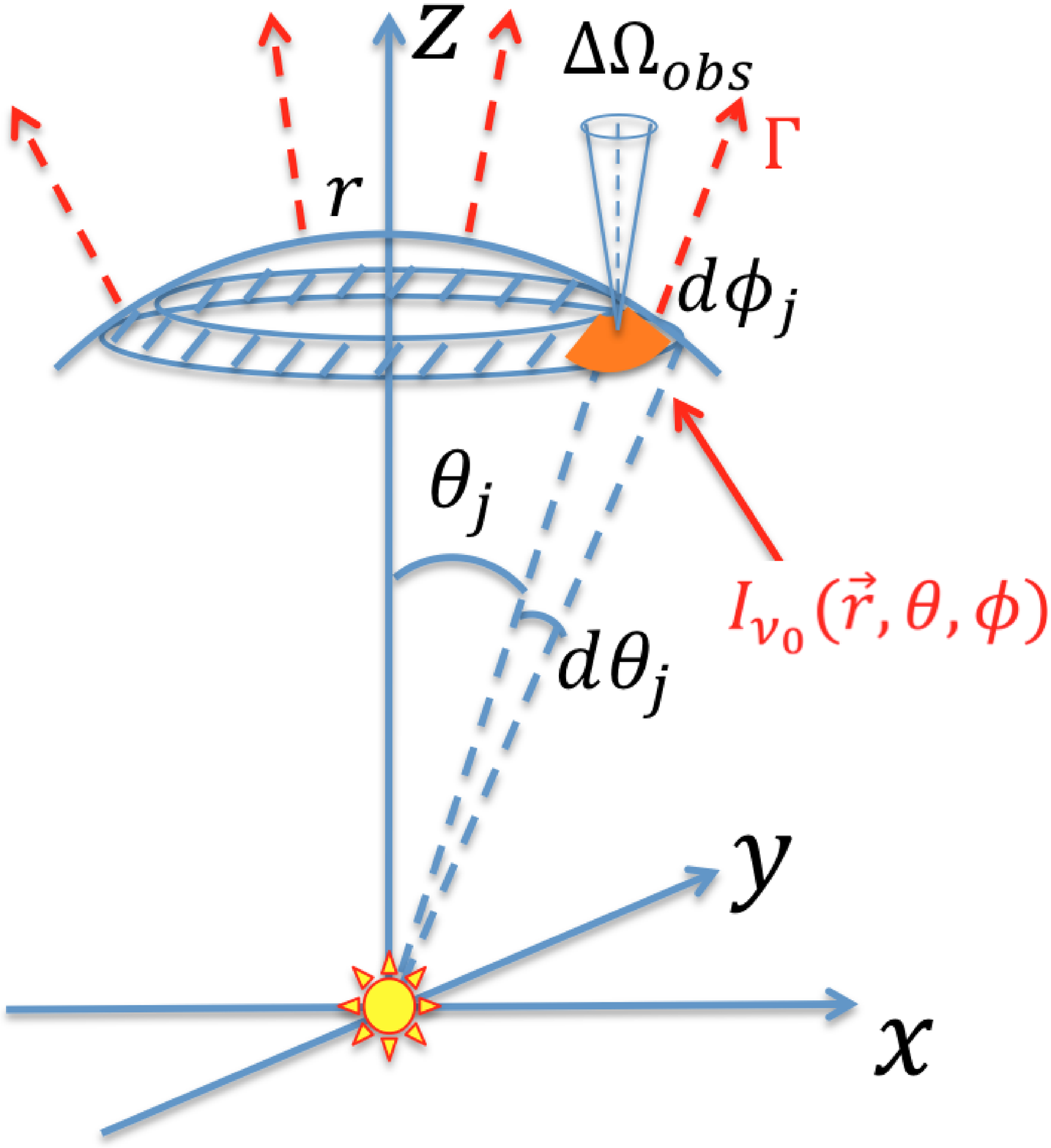}
  \caption{\small Lab frame (progenitor's rest frame). A spherically
    capped jet is moving in the z 
    direction with bulk Lorentz factor $\Gamma$. We consider a small
    volume element (\textcolor{orange}{orange}) $r^2d\Omega_j dr = r^2 
    sin\theta_j d\theta_j d\phi_j dr$ of the
    jet at position $\vec{r} = (r,\theta_j,\phi_j)$. The
    external photon field at position 
    $\vec{r}$ is denoted by intensity $I_{\nu_0} (\vec{r}, \theta,
    \phi)$, where $(\theta, \phi)$ is photons' moving
    direction. Photons scattered into the observer's cone $\Delta
    \Omega_{obs}$ are considered as observed.}
  \label{fig:labframe}
\end{figure}

As shown in Fig.(\ref{fig:labframe}),
the jet is moving with a Lorentz
factor (LF) $\Gamma$ in the z direction at radius $r$. For a volume
element (\textcolor{orange}{orange}) at position
$\vec{r}=(r,\theta_j,\phi_j)$, our goal is to 
consider all the EPF that is scattered
into the observer's cone $\Delta \Omega_{obs}$, which is a very small
solid angle in the z direction considering the cosmological
distances\footnote{When calculating the isotropic equivalent
  luminosity, we use $\Delta\Omega_{obs} = 4\pi$. Throughout this
  paper, if not specially stated, luminosities, emission energies,
  frequencies and observer's time are presented in the host galaxy
  frame.}. The EPF specific intensity is denoted as  
$I_{\nu_0} (\vec{r}, \theta, 
\phi)$, where $\vec{r}$ is the position of the volume element and
$(\theta,\phi)$ is the direction in which photons are moving.

In the comoving frame of
the volume element, the observer's cone is
\begin{equation}
\label{eq:18}
\Delta \Omega_{obs}^\prime =D_j^2 \Delta \Omega_{obs}
\end{equation}
where $D_j$ is the Doppler factor
\begin{equation}
  \label{eq:20}
  D_j = \frac{1}{\Gamma (1-\beta cos\theta_j)}
\end{equation}
and $\beta = \sqrt{1-1/\Gamma^2}$ is the velocity of the jet.

\begin{figure}
  \centering
  \includegraphics[width=0.35\textwidth,
  height=0.23\textheight]{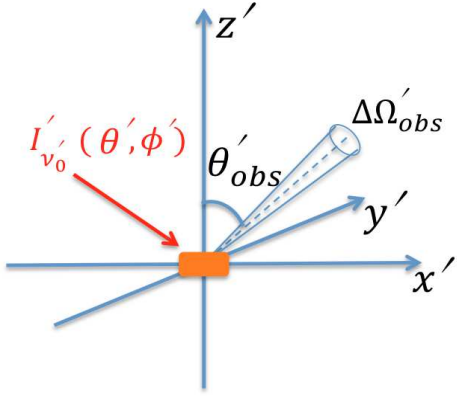}
  \caption{\small Comoving frame of the volume element
    (\textcolor{orange}{orange}) in 
    Fig.(\ref{fig:labframe}). External photon field is denoted by
    intensity $I_{\nu_0^\prime}^\prime(\theta^\prime,
    \phi^\prime)$. The observer's cone $\Delta 
    \Omega_{obs}^\prime$ is in the $x^\prime z^\prime$ plane. Photons
    scattered into $\Delta \Omega_{obs}^\prime$ 
  are considered as observed.}
  \label{fig:comovingframe}
\end{figure}

Fig.(\ref{fig:comovingframe}) shows the geometry in the comoving
frame, where $z^\prime$ axis is aligned with the original moving
direction of the volume element.
We orient the $x^\prime$ axis in a way that the 
observer's cone is in the $x^\prime z^\prime$ plane and at an angle
$\theta_{obs}^\prime$ with the $z^\prime$ axis. Recall that in the lab
frame, $\theta_j$ is the angle between observer's cone (same as z
axis) and the moving direction of the volume element. Therefore,
Lorentz transformation (LT) gives
\begin{equation}
  \label{eq:21}
  cos \theta_{obs}^\prime = \frac{cos \theta_j - \beta}{1-\beta cos\theta_j}
\end{equation}

Going from $I_{\nu_0} (\theta, \phi)$ to $I_{\nu_0^\prime}^\prime
(\theta^\prime, \phi^\prime)$ needs an axis rotation and a LT. The
axis rotation from  $xyz$ frame to
$\tilde{x}\tilde{y}\tilde{z}$ frame,
in which $\tilde{z}$ axis points towards the moving direction, can be
expressed by 
\begin{equation}
  \label{eq:22}
  (\vec{e_1}, \vec{e_2}, \vec{e_3}) = R_{\tilde{y}}(\theta_j) R_{\tilde{z}}(\pi - \phi_j)
  (\vec{\tilde{e_1}}, \vec{\tilde{ e_2}}, \vec{\tilde{ e_3}})
\end{equation}
where $\{\vec{e_i}\}$ and $\{\vec{\tilde{e_i}}\}$ ($i = 1,2,3$) are
respectively the basis
vectors in the $xyz$ and $\tilde{x}\tilde{y}\tilde{z}$
frame. $R_{\tilde{y},\tilde{z}}(\alpha)$ are rotation 
operators, along $\tilde{y}$ or $\tilde{z}$ axis by
an angle $\alpha$ according to the right-hand rule. Therefore, a
one-to-one map $(\theta,\phi)\rightarrow
(\tilde{\theta},\tilde{\phi})$ is obtained from Eq.(\ref{eq:22}).
Then LT gives the
intensity in the comoving frame
\begin{equation}
  \label{eq:25}
  I_{\nu_0^\prime}^\prime(\theta^\prime, \phi^\prime) =
  \tilde{I}_{\nu_0} (\ttheta, \tphi) \cdot \left(\frac{\nu_0^\prime}{\nu_0}\right)^3
\end{equation}
and frequency
\begin{equation}
  \label{eq:29}
  \nu_0 = \Gamma(1+\beta cos\theta^\prime)\nu_0^\prime
\end{equation}
and direction
\begin{equation}
  \label{eq:26}
 cos\ttheta = \frac{cos \thetap + \beta}{1+\beta cos\thetap},
\mbox{ } \tphi = \phip
\end{equation}

In the comoving ($x^\prime y^\prime z^\prime$) frame, if electrons are
assumed to be moving isotropically with LF
$\gamma_e$, the (average) differential IC cross section from
$(\nu_0^\prime, 
\thetap, \phip)$ to $(\nup,\theta_{obs}^\prime,\phi_{obs}^\prime = 0)$
is\footnote{The approximation is good if $\gamma_e\gg1$.}
\citep{1981Ap&SS..79..321A} 
\begin{equation}
  \label{eq:35}
  \frac{\partial^2\sigma}{\partial\nup\partial\Omega_{obs}^\prime} \simeq
  \frac{3\sigma_T}{16\pi\gamma_e^2 \nu_0^\prime} [1 +
  \frac{z^2}{2(1-z)} - \frac{2z}{b_\theta (1-z)} + \frac{2z^2}{b_\theta^2(1-z)^2}]
\end{equation}
where $z = h\nup/(\gamma_em_ec^2)$, $b_\theta =
2(1-cos\theta_{ic}^\prime)\gamma_e 
h\nu_0^\prime/(m_ec^2)$, $h\nu_0^\prime\ll h\nup\leq
\gamma_em_ec^2b_\theta /(1 + b_\theta)$, $\theta_{ic}^\prime$ is
the angle between the direction of incident and scattered
photons, i.e. $cos\theta_{ic}^\prime = sin\theta_{obs}^\prime
cos\phip sin\thetap + cos\theta_{obs}^\prime cos\thetap$, $m_e$
is electron mass, $c$ is speed of light, and $\sigma_T$ is Thomson
cross section.

We consider a powerlaw distribution of electrons
\begin{equation}
  \label{eq:32}
dn_e
  \begin{cases}
    \propto \gamma_e^{-p} d\gamma_e &\mbox{ if
    }\gamma_e\geq\gamma_m \\
= 0 &\mbox{ otherwise}
  \end{cases}
\end{equation}
Therefore, the averaged differential cross section becomes
\begin{equation}
  \label{eq:38}
  \begin{split}
    & \left<
      \frac{\partial^2\sigma}{\partial\nup\partial\Omega_{obs}^\prime}\right>_{\gamma_e}
    = 
  \frac{\int_{\gamma_m}^\infty
    \frac{\partial^2\sigma}{\partial\nup\partial\Omega_{obs}^\prime}
    \gamma_e^{-p}d\gamma_e}{\int_{\gamma_m}^\infty\gamma_e^{-p}d\gamma_e}\\
& =  (p-1) \left(\frac{\gamma_m m_ec^2}{h\nup}\right)^{p+1}
\frac{3\sigma_T}{16\pi \gamma_m^2\nu_0^\prime } \\
&\cdot \int_0^{z_{max}} \left[1 + \left(\frac{1}{2} - \frac{2}{a_\theta}\right)
\frac{z^2}{1-z} + \frac{2}{a_\theta^2}\frac{z^4}{(1-z)^2}\right]z^p dz
  \end{split}
\end{equation}
where $a_\theta =
2(1-cos\theta_{ic}^\prime)h^2\nu_0^\prime\nup/(m_ec^2)^2$, $z_{max} =
\mbox{min}(h\nup/(\gamma_m m_ec^2), (\sqrt{a_\theta^2 + 4a_\theta} -
a_\theta)/2)$. Note that $a_\theta\simeq h^2\nu_0^\prime
\nu^\prime/(m_ec^2)^2$ is roughly the criterion for whether EIC
scattering is in the Klein-Nishina (KN) regime. 

In the low energy band $\nup/(\gamma_m^2 \nu_0^\prime) <1$,
the differential cross section $\propto\nup^0$ (i.e. constant), so we
expect $\nup L_\nup\propto 
\nup^2$. In the high energy band $\nup/(\gamma_m^2 \nu_0^\prime) >1$,
we have 
\begin{equation}
  \label{eq:39}
  \begin{split}
    z_{max} &= \left(\sqrt{a_\theta^2 + 4a_\theta} -a_\theta\right)/2 \\
& =
    \begin{cases}
      1 - 1/a_\theta + O(1/a_\theta^2) & \mbox{ if } a_\theta \gg 1\\
\sqrt{a_\theta} + O(a_\theta) & \mbox{ if } a_\theta \ll 1
    \end{cases}
  \end{split}  
\end{equation}
Then we get
\begin{equation}
  \label{eq:40}
  \left<
      \frac{\partial^2\sigma}{\partial \nup \partial
        \Omega_{obs}^\prime}\right>_{\gamma_e} 
  \propto
  \begin{cases}
    \nup^{-(p+1)/2} & \mbox{ if } a_\theta \ll 1 \\
\nup^{-(p+1)} (ln(a_\theta) + C) & \mbox{ if } a_\theta \gg 1
  \end{cases}
\end{equation}
as pointed out by \citet{1970RvMP...42..237B} and
\citet{1981Ap&SS..79..321A}. Therefore, in the Thomson regime
($a_\theta \ll 1$), we expect $\nup L_\nup\propto \nup^{(3-p)/2}$; in
the KN regime ($a_\theta \gg 1$), we expect $\nup
L_\nup\propto \nup^{1-p}$. However, as we shall
see in Section \ref{lc_spectr}, the EIC spectrum is almost never
$\nu^{(3-p)/2}$ or $\nu^{1-p}$. The reasons are as follows: (1) When
$\nu/(\Gamma^2\gamma_m^2 \nu_0)<1$, we are generally in the Thomson
regime, and hence we get $\nu L_\nu\propto \nu^2$. (2) When 
$\nu/(\Gamma^2\gamma_m^2 \nu_0)>1$, we almost immediately get into the
KN regime ($a_\theta>1$), and the $(ln(a_\theta) + C)$ factor can't be
ignored until $a_\theta\geq100$ (i.e. $h\nu\geq 3\mbox{ }TeV$). As the
high energy tail suffers from pair-production absorption (see the
Appendix), the spectra hardly reach $TeV$ band.

Suppose the volume element (\textcolor{orange}{orange} in
Fig.\ref{fig:labframe}) contains 
$dN_e$ electrons, the  
number of photons that are scattered into $\Delta\Omega_{obs}^\prime$, in
a time duration $dt^\prime$ and frequency range $d\nup$, is
\begin{equation}
  \label{eq:41}
  \begin{split}
   & dN_\gamma = dN_e \cdot d\nup
  \Delta\Omega_{obs}^\prime dt^\prime\\
& \cdot\int_{4\pi}d\Omega^\prime\int_0^\infty d\nu_0^\prime \frac{
  I_{\nu_0^\prime}^\prime (\theta^\prime,
  \phi^\prime)}{h\nu_0^\prime} \left<
  \frac{\partial^2\sigma}{\partial\nup\partial
    \Omega_{obs}^\prime}\right>_{\gamma_e} 
e^{-\tau(\nu_0^\prime, \thetap)}  
  \end{split}
\end{equation}
where the optical depth $\tau(\nu_0^\prime, \thetap)$ describes the
attenuation by the part 
of the jet that lies along the incident photons' trajectory before they
enter the jet volume element where the IC scattering is
considered. Second scattering is ignored here. We assume the jet is
launched steadily with (isotropic) 
power $L_j$ and duration $T_j$, total (isotropic) kinetic
energy is $E_j$, and the thickness of the jet is $\Delta r = c
T_j$. Photons are all assumed to enter the jet from the front
surface, which is at radius $r$. If the volume element is located
$\delta r$ from the jet 
front, the optical depth along the incident photons' trajectory is
\begin{equation}
  \label{eq:45}
  \tau(\nu_0^\prime, \thetap) = \frac{\delta r}{\Delta r} \frac{E_j}{\Gamma m_p c^2}
  \sigma_{tot}(\nu_0^\prime) \cdot \frac{1}{4\pi r^2 |cos\thetap|}
\end{equation}
where $|cos\thetap|$ is due to the incident angle onto the front
surface and $\sigma_{tot} (\nu_0^\prime)$ is the total KN cross
section averaged over electrons' powerlaw distribution
\begin{equation}
  \label{eq:46}
  \sigma_{tot} (\nu_0^\prime) = \frac{\int_{\gamma_m}^\infty
    \sigma_{KN}(\nu_0^\prime, \gamma_e)
    \gamma_e^{-p}d\gamma_e}{\int_{\gamma_m}^\infty\gamma_e^{-p}d\gamma_e}
\end{equation}

Actually, not all the $dN_\gamma$ (Eq.\ref{eq:41}) photons will be
observed. Photons of energy $h\nu\gtrsim 10\mbox{ }GeV$ may collide
with low energy photons (such as the GRB prompt 
emission) and produce pairs. In the Appendix, we calculate the
pair-production optical depth $\tau_{\gamma\gamma} (\nu, r)$
(Eq.\ref{eq:59}) for high energy photons produced at radius
$r$ and have frequency 
$\nu$. Basically, we show that only the absorption by the GRB prompt 
emission may be important, if the prompt emission is produced at the
same radius. And we use a broken
powerlaw spectrum \citep{1993ApJ...413..281B} with low (high)
frequency index $-1$ ($-2.4$), break frequency $200 \mbox{ } keV$ and
bolometric isotropic luminosity $10^{52} \mbox{ } erg/s$, following the
{\it Fermi} GRB statistics \citep{2014ApJS..211...12G}. Then, the
$dN_\gamma$ in Eq.(\ref{eq:41}) needs to be multiplied by
$e^{-\tau_{\gamma\gamma}}$. 

Next we include the effect of the conical shape of the jet (``curvature
effect'') on the observed luminosity.
When the jet front is at radius $r$, the scattered photons from a
volume element at a distance $\delta r$ from the jet front and 
at latitude $\theta_j$ will arrive at observer's time $t_{obs} (r, \delta
r, \theta_j)$. Photons arriving at the same time are from the same
equal-arrival time surface, described by
\begin{equation}
  \label{eq:48}
     r_{eq}(\theta_j,\delta r,t_{obs}) = \beta\frac{ct_{obs} -
     \delta r}{1-\beta cos\theta_j}
\end{equation}
We divide the jet into many thin shells of thickness
$d(\delta r)$ according to the distance to the jet front $\delta r$
and calculate the lightcurve contributed by each shell.
Using Lorentz transformation in time
$dt^\prime = dr_{eq}/(\Gamma c) \simeq D_j dt_{obs}$ and frequency
$\nup = \nu/D_j$ in Eq.(\ref{eq:41}), we obtain the specific
luminosity from one shell $d(\delta r)$ at $\delta r$
\begin{equation}
  \label{eq:50}
  \begin{split}
    &  dL_\nu = \frac{dN_\gamma h\nu}{d\nu dt_{obs}} e^{-\tau_{\ggam}(\nu,r)} \\
&= \frac{E_j}{\Gamma m_p c^2}\frac{d(\delta r)}{\Delta r} \frac{\int
  d\Omega_j}{4\pi} h\nu D_j^2 \Delta 
  \Omega_{obs}  e^{-\tau_{\ggam}(\nu,r)} \\ 
& \cdot\int_{4\pi}d\Omega^\prime\int_0^\infty d\nu_0^\prime \frac{
  I_{\nu_0^\prime}^\prime(\theta^\prime, 
  \phi^\prime)}{h\nu_0^\prime} \left<
      \frac{\partial^2\sigma}{\partial\nup\partial
        \Omega_{obs}^\prime}\right>_{\gamma_e} e^{-\tau(\thetap)} 
  \end{split}
\end{equation}
where $\int d\Omega_j = \int_0^{\theta_{j,max}}sin\theta_j d\theta_j
\int_0^{2\pi}d\phi_j$ and we use $\theta_{j,max} = 4/\Gamma$ (not sensitive) in our
numerical results (Section \ref{lc_spectr}). Then, the total observed
luminosity is obtained by adding up all the shells, i.e. integrating
Eq.(\ref{eq:50}) over $\int_0^{\Delta r}d(\delta r)$.

\section{Applications}
In this section, we apply the procedures developed in Section
\ref{section:theory} to three different cases, where the EPFs are
given by the physical environment. First, in Section
\ref{section:oom}, we give simple 
order-of-magnitude estimates of the total EIC energy, not considering
the pair production. See Table (\ref{tab:allcases}) for the summary of our
estimates on peak frequency, EIC luminosity, duration, etc in all cases.
Then in Section \ref{lc_spectr}, we present the EIC lightcurves and
spectra, with absorption from pair production (with prompt $\gamma$-rays)
considered. For simplicity, each case is considered
  separately in this paper, but in reality, different cases can
  operate simultaneously (e.g. the GRB progenitor is in a binary
  system and has a wind) and the total EIC emission will approximately
  be a superposition of them.

\subsection{Order-of-magnitude Estimate}\label{section:oom}
\subsubsection{Case (I) --- Binary System}
Consider the GRB progenitor star being in a binary system. Soft photons
from the companion star are scattered by the electrons
in the relativistic jet.

\begin{figure}
  \centering
  \includegraphics[width=0.35\textwidth,
  height=0.23\textheight]{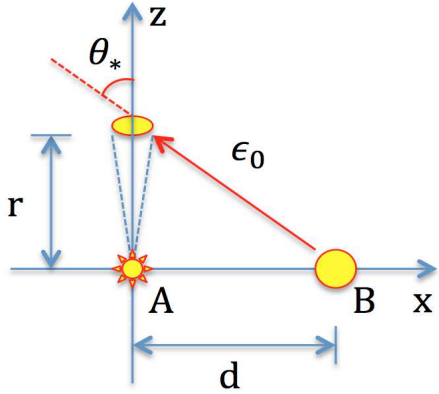}
  \caption{\small Geometry of {\it Case (I)}. The jet is on the z axis
    at a distance $r$ from the progenitor star (A). The companion star
    (B) is on the x axis at a distance $d$ from Star A. Photons from
    Star B encounter the jet at an angle $cos\theta_* =
    r/\sqrt{r^2+d^2}$.}
  \label{fig:case1}
\end{figure}

We assume the direction of the jet is perpendicular to the orbital
plane\footnote{Even if the jet direction is inclined by up to
  $30^o$ from the normal direction of orbital plane, the total energy (or
  luminosity) will be affected by a factor of order unity ($\lesssim2$).}. As
shown in Fig.(\ref{fig:case1}),  
the jet is at a distance $r$ from the progenitor star (A) and is
moving with LF $\Gamma$ in the z direction, and the
companion star (B) is on the x axis at a distance $d$ from A. Photons
from Star B encounter the jet at an angle $cos\theta_* = r/\sqrt{r^2+d^2}$.

Suppose the companion star has a bolometric luminosity $L_b =
10^{39}erg/s\mbox{ }L_{b,39}$ and peak
energy $\epsilon_0 = 10eV\mbox{ }\epsilon_{0,1}$.
Considering the possible recoil of electrons, we 
estimate the energy of scattered photons by
\begin{equation}
  \label{eq:51}
  \begin{split}
     \epsilon &\simeq \mbox{min}\left(\Gamma^2\gamma_e^2
  \epsilon_0,\Gamma\gamma_e m_ec^2\right)\\
&= 9GeV\mbox{ }
  \Gamma_{2.5}^2\gamma_{e,2}^2 \epsilon_{0,1}\eta_{KN}
  \end{split}
\end{equation}
where we have used fiducial LFs $\Gamma = 300 \Gamma_{2.5}$,
$\gamma_e = 10^2 \gamma_{e,2}$, and
\begin{equation}
  \label{eq:61}
  \eta_{KN} = \mbox{min}\left(1,
    \frac{1.7}{\Gamma_{2.5}\gamma_{e,2} \epsilon_{0,1}}\right)
\end{equation}
describes the degree to which EIC scattering is KN
suppressed. Suppose 
the jet is launched steadily with (isotropic) power $L_j =
10^{53}erg/s\mbox{ }L_{j,53}$ and
duration $T_j = 10s\mbox{ }T_{j,1}$, so the total (isotropic) kinetic
energy is $E_j = 10^{54}erg\mbox{ }E_{j,54}$ and the 
thickness of the jet is $\Delta r = c T_j = 3\times10^{11}cm\mbox{
}T_{j,1}$. Considering possible KN suppression, we estimate the EIC
optical depth of the whole jet as
\begin{equation}
  \label{eq:1}
  \begin{split}
    \tau_{j}(r)\simeq \frac{E_j \sigma_T}{4\pi  r^2 \Gamma m_p c^2 } 
\eta_{KN}
  \end{split}
\end{equation}
The jet becomes transparent to external photons at radius
\begin{equation}
  \label{eq:5}
 r_{tr}\simeq 3.4\times10^{14}cm\mbox{ }
\sqrt{\frac{E_{j,54}}{\Gamma_{2.5}} \eta_{KN}}
\end{equation}
Here, to estimate the total number of EIC scattered
  photons, we consider external photons that are in the $\pi
  \theta_{j, max}^2$ solid angle to be swept by the jet
  ($\theta_{j, max}$ being the half opening angle of the jet). As the
  jet sweeps through the volume $\pi \theta_{j, max}^2 r^2 dr$, the
  probabiliy of an external photon being scattered by the jet is
  $min(1, r_{tr}^2/r^2)$. This is
  because: below $r_{tr}$, the jet is optically thick and all the
  external photons contribute to the EIC flux; above $r_{tr}$, the
  effective cross section of the jet is the sum of individual
  electrons and that equals to $\pi \theta_{j, max}^2 r_{tr}^2$ and is
  smaller than $\pi \theta_{j, max}^2 r^2$. Note that $r_{tr}$ in
  Eq.(\ref{eq:5}) is different from the photospheric radius in that
  the former considers all electrons in the jet but the latter only
  considers electrons in the casualty connected thickness
  $r/2\Gamma^2$ (usually $\ll cT_j$). We note the fact
  that EIC flux from outside the casualty connected region (deeper 
  layers in the jet) will arrive at the observer at later time, and
  this is fully taken into account when we calculate the spectra and
  lightcurves (through Eq. \ref{eq:48}).

The number density of external
  soft photons is given by
\begin{equation}
  \label{eq:54}
  n_\gamma = \frac{L_b}{4\pi (r^2 + d^2)\epsilon_0 c}
\end{equation}

The total (isotropic) number of scattered photons is
\begin{equation}
  \label{eq:53}
      N_\gamma = \int_0^\infty (1-\beta cos\theta_*) n_\gamma(r,d)
      \mbox{min}(1, \tau_j(r))
      4\pi r^2 dr
\end{equation}
from which, we can easily see that most photons are scattered at
radius $r \sim d$. Therefore, we can estimate
\begin{equation}
  \label{eq:55}
  N_\gamma \sim 3\times 10^{53}\mbox{ }
  \frac{L_{b,39}d_{15}}{\epsilon_{0,1}}\mbox{min}\left(1,\frac{r_{tr}^2}{d^2}\right)
\end{equation}
Multiplying $N_\gamma$ by the scattered photons' energy from
Eq.(\ref{eq:51}), we get the total EIC energy 
\begin{equation}
  \label{eq:56}
  \begin{split}
      E_{EIC}\sim 4\times10^{51} erg\mbox{
      }\Gamma_{2.5}^2\gamma_{e,2}^2 \eta_{KN} L_{b,39}d_{15} 
\mbox{min}\left(1, \frac{r_{tr}^2}{d^2}\right) 
  \end{split}
\end{equation}
Note that the estimate is not accurate when $r_{tr}\gg d$, because the soft
photons are moving nearly parallel to jet, and hence both $N_\gamma$ and
$\epsilon$ decrease by orders of magnitude.

\subsubsection{Case (II) --- Stellar Wind}
\begin{figure}
  \centering
\includegraphics[width=0.35\textwidth,
  height=0.23\textheight]{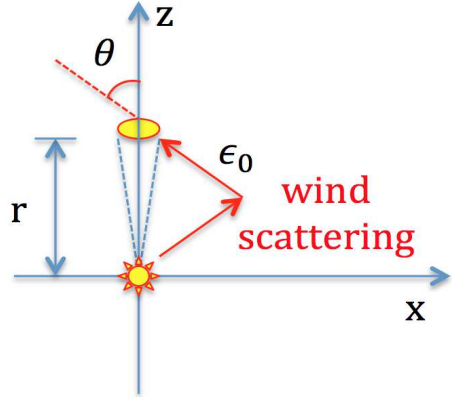}
  \caption{\small Geometry of {\it Case (II)}. The jet is on the z axis at a distance $r$
from the progenitor star. Photons originally from Star A are scattered first
by electrons in the wind and again by the jet. The wind scattered
photons encounter the jet with an angle $\theta$.}
  \label{fig:case2}
\end{figure}

As suggested by e.g. \citet{1993ApJ...405..273W} and
\citet{1999ApJ...524..262M}, GRB progenitor
stars are possibly Wolf-Rayet (W-R) stars, which have a typical mass loss
rate of $10^{-5}M_{\odot}/yr$, bolometric
luminosity $10^{39}\mbox{ }erg/s$, and effective temperature
$10^5\mbox{ }K$ \citep{2007ARA&A..45..177C}. We
consider that photons originally from the progenitor star are scattered first by
the wind and then by the jet. The geometry is shown in Fig.(\ref{fig:case2}).

We assume a steady wind and denote the wind velocity\footnote{The
  order of escape velocity.} as $v =10^8 cm/s \mbox{ } v_8$ and
the mass loss rate as $\dot{M}
=10^{-5}M_{\odot} /yr\mbox{ }{\dot{M}_{-5}}$. The wind-scattering optical
depth of the region with radius $>r$ is
\begin{equation}
  \label{eq:27}
  \begin{split}
    \tau_{wind}(r) &= \sigma_T \int_r^\infty \frac{\dot{M}}{4\pi \tilde{r}^2 v
    m_p c^2} d\tilde{r}\\
&\simeq 2\times 10^{-3} \frac{\dot{M}_{-5}}{v_8}r_{14}^{-1}
  \end{split}
\end{equation}
Therefore, the wind-scattered photon density is
\begin{equation}
  \label{eq:42}
  n_\gamma(r) = \frac{L_b \tau_{wind}(r)}{4\pi r^2 \epsilon_0 c}
  (\propto r^{-3})
\end{equation}
The number of scattered photons has a logarithmic dependence on $r$
when $r<r_{tr}$, and then drops off as $r^{-2}$ at larger radius, so
we can estimate the total (isotropic) number of scattered photons as
\begin{equation}
  \label{eq:43}
  N_\gamma = 2.9\times 10^{51} \mbox{ }\frac{L_{b,39}
    \dot{M}_{-5}}{\epsilon_{0,1}v_8} ln \left(\frac{r_{tr,14}}{R_{*,11}}\right)
\end{equation}
where $R_{*} = 10^{11}cm \mbox{ }R_{*,11}$ is the radius of the star.
Therefore, the total EIC energy is
\begin{equation}
  \label{eq:44}
  \begin{split}
      E_{EIC} = 4.2\times 10^{49} erg \mbox{ } \frac{L_{b,39}
    \dot{M}_{-5}\Gamma_{2.5}^2 \gamma_{e,2}^2\eta_{KN}}{v_8} ln
  \left(\frac{r_{tr,14}}{R_{*,11}}\right) 
  \end{split}
\end{equation}

\begin{table*}
 \centering
 \begin{minipage}{\textwidth}
  \caption{Summary of the EIC emission from our analytical
  calculation (Section \ref{section:oom}, not considering
  pair-production absorption and electrons having a single LF),
  including $h\nu_p$ ({\bf \small peak 
  energy}), $L_{EIC}^{iso}$ ({\bf \small isotropic EIC luminosity}), $t_{obs}$ 
  ({\bf \small duration of EIC luminosity}), 
  $N_{EIC}^{iso}$ ({\bf \small isotropic number of EIC photons at
    $\sim\nu_p$}).
  The scaled parameters are all set to 1, including $E_{j,54}$ (isotropic jet
  kinetic energy), $T_{j,1}$ (jet 
  duration), $\Gamma_{2.5}$ (jet bulk LF), $\gamma_{e, ID, 2}$
  (ID-accelerated electrons' LF), $r_{max,16}$ (the maximum radius
  electrons stay hot), $\Gamma_{dec,2}$ (bulk LFs of
  FS and RS at deceleration radius), $\gamma_{e,RS, 2.5}$
  (RS-accelerated electrons' LF), $r_{dec,17}$ (deceleration radius),
  $T_{4.7}$ (temperature of EPF-contributing stars), $L_{b,39}$
  (bolometric luminosity of EPF-contributing stars), $\dot{M}_{-5}$
  (wind mass loss rate), $v_{8}$ (wind velocity), $n_{*,4}$ (stellar
  number density in the cluster), $R_{pc}$ (radius of the cluster).}
  \label{tab:allcases}
  \begin{tabular}{@{}cccccccc@{}}
  \hline\hline
   Cases    &     {\it I ($d=10^{14}$)}  & {\it I ($d=10^{15}$)} & {\it I ($d=10^{16}$)}  
        & {\it II}  & {\it III (ID)}   & {\it
  III (RS)}  &  {\it III (FS)} \\
 \hline
$h\nu_p$ [$GeV$]
& 9  & 9 & 9  & 9 & 9 & 9 & 26\\[0.3cm]

$L_{EIC}^{iso}$ [$erg/s$]
& $4\times10^{49}$  & $5\times10^{49}$ & $5\times10^{48}$  & $5\times10^{48}$
& $10^{49}$ & $5\times10^{48}$ & $2\times10^{46}$\\[0.3cm]

$t_{obs}$ [$s$]
& 10  & 10 & 10  & 10 & 10 & $2\times10^2$ & $2\times10^4$\\[0.3cm]

$N_{EIC}^{iso}$\footnote{$10^{52}$
  photons from $1\mbox{ }Gpc$ ($z=0.25$) give a photon fluence
  of $10^{-4}\mbox{ }cm^{-2}$. With a collecting area $\sim 10^4\ cm^2$,
    {\it Fermi} Large Area Telescope (LAT) is capable of probing the
    EPF around low-redshift 
    ($z\lesssim 0.5$) GRB progenitor stars. Also note that he EIC
    photons from the forward shock in {\it Case (III)} span a
    large range of energies. The isotropic number of EIC photons in
    $0.1-300\mbox{ }GeV$ ({\it Fermi} LAT sensitivity range) is
    $N_{EIC}^{iso}\simeq 2.5\times 10^{53}$ and the duration is
    $t_{obs}\simeq 9\ d$ (see Section \ref{model_star_cluster}).}
& $3\times10^{52}$  & $3\times10^{52}$ & $3\times10^{51}$ 
& $3\times10^{51}$  &$10^{52}$ & $10^{53}$ & $6\times10^{51}$\\[0.3cm]

$r_{eff}$\footnote{By using a
  constant $\gamma_e$ at all radius in
  our analytical and numerical 
  calculations, we are making an assumption that 
  electrons are hot near the radius $r_{eff}$ where most scatterings
  happen (see Fig.\ref{fig:case1_rlimited} for a justification).} [$cm$]
& $d(10^{14})$  & $d(10^{15})$ & $d(10^{16})$ & $r_{tr}
(3\times10^{14})$ & $r_{max} (10^{16})$ 
& $r_{dec}(10^{17})$ & $r_{KN}(3.3\times10^{17})$\\
\hline\hline
\end{tabular}
\end{minipage}
\end{table*}

\subsubsection{Case (III) --- Star Cluster}\label{model_star_cluster}
Consider the GRB progenitor star being in a young massive
cluster. All the stars in the cluster create a nearly isotropic 
and uniform EPF around the GRB jet. Scattering by
electrons accelerated by both internal dissipation (ID) and external shocks
(ES) may be important. We assume that the cluster has a(n) (O-)star
density of $n_* = 10^4 pc^{-3}\mbox{ } n_{*,4}$ and a radius of
$R = 1 pc \mbox{ }R_{pc}$. We assume stars have an average bolometric
luminosity of $L_b = 10^{39}erg/s \mbox{ }L_{b,39}$ and peak energy
$\epsilon_0 = 10 eV\mbox{ }\epsilon_{0,1}$. The
number density of EPF can be estimated as
\begin{equation}
  \label{eq:52}
  n_\gamma \simeq \frac{n_* L_b R}{\epsilon_0 c} \simeq 2\times10^6
  cm^{-3}\mbox{ } \frac{n_{*,4}L_{b,39} R_{pc}}{\epsilon_{0,1}}
\end{equation}

First, we consider ID. After the jet becoming transparent
($r>r_{tr}$), $4\pi r^2n_\gamma\tau_j$ is a constant. The total number
of scattered photons depends on the largest radius where electrons
stay hot $r_{max}$, which is quite uncertain \citep{2004RvMP...76.1143P,
2006RPPh...69.2259M, 2015PhR...561....1K}. Observationally, the
duration of GRB prompt emission ($T_{90}/(1+z)\sim 10\mbox{ }s$) gives
an upper limit of the radius where the ID is active
\begin{equation}
  \label{eq:114}
  r_{ID}\leq 2\Gamma^2 c T_{90}/(1+z)\simeq 5\times10^{16}cm \mbox{ }
\Gamma_{2.5}^2
\end{equation}
Theoretically, an upper limit of $r_{ID}$ is the deceleration radius
$r_{dec}$ (see Eq.\ref{eq:9} below), where the jet starts to
decelerate substantially. Note that,
after the prompt emission is produced, electrons cool via adiabatic
expansion $\gamma_e \propto r^{-2/3}$ and may 
stay hot for some time. Putting the
uncertainties in the parameter $r_{max} = 10^{16}cm\mbox{
}r_{max,16}$, we estimate the total EIC energy by
\begin{equation}
  \label{eq:161}
   \begin{split}
     E_{EIC}\simeq & (\epsilon \tau_j n_\gamma 4\pi r^3/3)|_{r = r_{max}}\\
\simeq & 1.4\times 10^{50} erg\mbox{ }
E_{j,54}\Gamma_{2.5}\gamma_{e,2}^2\eta_{KN}^2\\ 
 &\cdot n_{*,4}L_{b,39}R_{pc}\cdot r_{max,16}
  \end{split}
\end{equation}

Next, we consider the EIC emission from ES, where both the
reverse shocked (RS) and forward shocked (FS) region may contribute
significantly. EIC emission from ES strongly depends on the ES dynamical
evolution.

We assume pressure equilibrium\footnote{This may only hold
  for a so-called short-lived RS, but see
  \citet{2011ApJ...733...86U} for a description of a long-lived RS,
  where RS may reach a much larger radius. Then the EIC energy from RS
 could be much larger.}
between FS and RS region \citep{2000ApJ...545..807K}, both moving
together with LF $\Gamma_{sh}$. Then 
$\Gamma_{sh}$ is only a function
of jet LF $\Gamma$ and the density ratio $n/n_{ej}$, where $n$
is the number density of the circum-burst medium and $n_{ej}$ is
the comoving number density of unshocked ejecta. Before
RS crosses the jet, the LF of the shocked region is
\begin{equation}
  \label{eq:121}
  \begin{split}
      \Gamma_{sh}(r)= \frac{\Gamma}{(1 +
      2\Gamma\sqrt{n(r)/n_{ej}(r)})^{1/2}}
  \end{split}
\end{equation}
The comoving number density of the unshocked ejecta is
\begin{equation}
  \label{eq:123}
  n_{ej}^\prime(r) = \frac{E_j}{4\pi r^2 T_j \Gamma^2 m_p c^3}
\end{equation}
The circum-burst medium is either a uniform density medium or
stratified like a wind ($n\propto r^{-2}$) out to the wind termination
shock radius. GRB afterglow analyses suggest
that about half of the long GRBs have a uniform or weak (1/100 W-R)
wind density profile, rather than a normal W-R wind profile
\citep[e.g.][]{2001ApJ...560L..49P, 2004ApJ...606..369C}, the reason
for which is still an open question. 

For simplicity, we take the
electrons' number density in the cluster to be uniform
$n=10cm^{-3}\mbox{ }n_1$. Then the jet dynamics is characterized by the 
deceleration radius $r_{dec}$ \citep[see
e.g.][]{2004RvMP...76.1143P}. At radius
$r<r_{dec}$, whether RS is relativistic ($\Gamma_{sh}\ll\Gamma$) or
Newtonian ($\Gamma_{sh}\approx\Gamma$) depends on the factor
in Eq.(\ref{eq:121}) $2\Gamma(n/n_{ej})^{1/2}\simeq
0.48\Gamma_2^2r_{17}(n_1T_{j,1} /E_{j,54})^{1/2}$.
For a fast jet, say $\Gamma \geq 10^3$, from Eq.(\ref{eq:121}), we have
\begin{equation}
  \label{eq:2}
  \Gamma_{sh}(r) \simeq 1.4\times10^2r_{17}^{-1/2}
  \left(\frac{E_{j,54}}{n_1 T_{j,1}}\right)^{1/4}\mbox{ }(\ll \Gamma)
\end{equation}
For a slow jet, say $\Gamma = 100$, $2\Gamma(n/n_{ej})^{1/2}\ll 1$, so
we have $\Gamma_{sh}(r\leq r_{dec})\simeq 100$.

The deceleration radius can be estimated by
\begin{equation}
  \label{eq:6}
  \Gamma_{sh}^2\frac{4\pi}{3}r_{dec}^3nm_pc^2 = E_j/2
\end{equation}
which gives
\begin{equation}
  \label{eq:9}
  r_{dec}\simeq 9.3\times 10^{16} cm \mbox{ }
  \frac{E_{j,54}^{1/3}}{\Gamma_{sh,2}^{2/3}n_1^{1/3}}
\end{equation}
Therefore, for a jet LF $\Gamma\in(10^2,10^3)$, we always have
$r_{dec}\sim 10^{17}\mbox{ } cm$ and $\Gamma_{dec} \equiv
\Gamma_{sh}(r_{dec}) \sim 100$. Hereafter 
we use $r_{dec} = 10^{17}cm\mbox{ }r_{dec,17}$ and $\Gamma_{dec} =
10^2\Gamma_{dec,2}$ as the fiducial values. 

Electrons in the RS and FS region are accelerated to different LFs.
Assuming a fraction $\epsilon_e$ of the shocked fluid's internal
energy goes into electrons, we have 
\begin{equation}
  \label{eq:135}
  \gamma_{e} \simeq
  \begin{cases}
    \epsilon_{e,r} \frac{1}{2}\left( \frac{\Gamma}{\Gamma_{sh}} +
      \frac{\Gamma_{sh}}{\Gamma} \right)  \frac{m_p}{m_e} &\mbox{ (RS)}\\ 
\epsilon_{e,f} \Gamma_{sh}
      \frac{m_p}{m_e} &\mbox{ (FS)}
  \end{cases}
\end{equation}
where $m_p$ and $m_e$ are proton and electron mass. Below, we consider
RS and FS separately.

{\it (1) RS}: EIC emission from RS peaks when RS crosses the end of
the jet. After shock crossing, a rarefaction wave propagates
through the RS region at sound speed and decreases the internal energy
substantially. Despite the difference between shock crossing time
(slightly earlier, typically) and deceleration time (later), a good
approximation is that RS region stays hot until $r_{dec}$. The
energy equipartition parameter $\epsilon_{e,r}$ is poorly constrained
from GRB afterglows, since there are not many confirmed RS
detections. Considering the uncertainties, we
use $\epsilon_{e,r} = 0.1$ (the same as in FS). Therefore, from
Eq.(\ref{eq:135}), we get $\gamma_e \simeq 1.8\times10^2 
(1 + \Gamma/2\Gamma_{sh})$, i.e. typically a few hundred. Hereafter,
we use $\gamma_e = 300\gamma_{e,2.5}$ as the fiducial value for RS.

According to Eq.(\ref{eq:51}), the scattered photons'
energy is $\epsilon\simeq 9 GeV\mbox{
}\Gamma_{dec,2}^2\gamma_{e,2.5}^2\epsilon_{0,1}$ (marginally KN
suppressed) and the total (isotropic) EIC energy from RS can
be estimated as 
\begin{equation}
  \label{eq:17}
  \begin{split}
     & E_{EIC}^{RS} = (\epsilon \tau_j n_\gamma 4\pi r^3/3)|_{r = r_{dec}}\\
& \simeq  1.4\times 10^{51}erg\mbox{
}\frac{E_{j,54}}{\Gamma_{2.5}}n_{*,4} L_{b,39}
R_{pc}\Gamma_{dec,2}^2\gamma_{e,2.5}^2r_{dec,17}
  \end{split}
\end{equation}

{\it (2) FS}: We use Eq.(\ref{eq:135}) with a constant $\epsilon_e= 0.1$
\citep{2001ApJ...560L..49P} to get the electrons' LF $\gamma_e$.  At $r_{dec}$, EIC
radiation from FS region is strongly KN suppressed, so the
total EIC energy is rising with $r$ ($E_{EIC}^{FS}\propto r^4$ as
shown below), until the scattering changes from KN to Thomson
regime. The characteristic radius where this transition happens is
denoted as $r_{KN}$, given by 
\begin{equation}
  \label{eq:28}
  \Gamma_{sh}(r_{KN})\gamma_e\epsilon_0 = m_ec^2
\end{equation}
After $r_{dec}$, we expect the Blandford-Mckee evolution 
\citep{1976PhFl...19.1130B}
\begin{equation}
  \label{eq:156}
  \Gamma_{sh}(r) \simeq \Gamma_{dec}
  \left(\frac{r}{r_{dec}}\right)^{-3/2}
\end{equation}
Putting Eq.(\ref{eq:135}) and (\ref{eq:156}) to Eq.(\ref{eq:28}), we
get the ``KN radius''
\begin{equation}
  \label{eq:30}
  r_{KN} \simeq 3.3 r_{dec}\Gamma_{dec,2}^{2/3}\epsilon_{0,1}^{1/3}
\end{equation}

We know $ E_{EIC}^{FS} \propto \epsilon \tau_{FS} r^3$. When
$r_{dec}<r<r_{KN}$, the scattered photons' energy $\epsilon\propto
r^{-3}$, and the optical depth of FS region $\tau_{FS}\propto r^4$, so
we get $E_{EIC}^{FS}\propto r^4$. In a similar way, when $r>r_{KN}$,
we get $E_{EIC}^{FS}\propto r^{-2}$. Therefore, most EIC energy is
produced at radius $r_{KN}$, and we obtain
\begin{equation}
  \label{eq:34}
  \begin{split}
      E_{EIC}^{FS} &\simeq (\epsilon\tau_{FS} n_\gamma
      4\pi r^3/3)|_{r = r_{KN}}\\
&\simeq 2.6\times10^{50}erg \mbox{ }\frac{n_{*,4} L_{b,39}
  R_{pc}}{\epsilon_{0,1}^{2/3}}n_1 r_{dec,17}^4 \Gamma_{dec,2}^{8/3}
  \end{split}
\end{equation}
where we have used $\epsilon(r_{KN}) \simeq
(m_ec^2)^2/\epsilon_0=26GeV\mbox{ }\epsilon_{0,1}^{-1}$
and $\tau_{FS}(r_{KN}) \simeq r_{KN}n\sigma_T/3$. We note
the strong dependence on both $r_{dec}$ and $\Gamma_{dec}$. 
Also, higher energy (up to $\sim TeV$) EIC photons are produced at radius
$r_{dec}<r<r_{KN}$, but they contribute a smaller EIC energy. At
radius $r>r_{KN}$, the decreasing of EIC energy $E_{EIC}^{FS}\propto
r^{-2}$ is mostly due to the fast decreasing of EIC photons' energy
$\epsilon \propto \Gamma^4\propto r^{-6}$. On the other hand, the total
isotropic number of EIC photons is rising with radius $N_{EIC}^{iso} = 4\pi
r^3\tau_{FS} n_{\gamma}/3\propto r^4$ and so as the observational
duration $t_{obs}\propto r/\Gamma^2\propto r^4$. Therefore, if we
use the the same fiducial parameters as in Eq.(\ref{eq:34}), (1) the
total isotropic number of 
photons in the $0.1-300\mbox{ }GeV$ energy range ({\it Fermi} Large Area
Telescope sensitivity) is $N_{EIC}^{iso}(r_{KN})
(26/0.1)^{4/6}\simeq 2.5\times 10^{53}$; (2) the observational
duration is $t_{obs}\simeq (26/0.1)^{4/6}
r_{KN}/[2\Gamma_{sh}^2(r_{KN})c]\simeq 9 \mbox{
}d$ and the photon number flux is constant with
time. \citet{2011MNRAS.418..583M} also calculated the EIC 
scattering of diffuse EPF in a dense star cluster by the forward
shock. They used a lower stellar number density and a more realistic
circum-burst medium density profile (by simulating 
the wind-wind collision between a Wolf-Rayet star and an O star), and
the results in this paper are consistent with theirs.

\subsection{Lightcurves and Spectra}\label{lc_spectr}
In this subsection, we put realistic EPF profiles into the procedure
developed in Section \ref{section:theory} and calculate the EIC
lightcurves and spectra for the three cases. As a general assumption,
in each case, 
the star(s) that contribute to EPF are assumed to have bolometric
luminosity $L_b  
=10^{39}erg/s\mbox{ }L_{b,39}$ and effective
temperature $T = 5\times 10^4K\mbox{ }T_{4.7}$, so the specific
luminosity is 
\begin{equation}
  \label{eq:31}
  L_{\nu_0} = \frac{\pi L_b}{\sigma_{SB} T^4}\cdot
  \frac{2h\nu_0^3}{c^2}\frac{1}{e^{h\nu_0/kT} - 1}
\end{equation}
where $\sigma_{SB}$ is the Stefan-Boltzmann constant.

{\it (I) Binary Case}: We consider the companion star
as a point source, so 
the EPF intensity at position $\vec{r} = (r,\theta_j,\phi_j)$, in the direction
($\theta,\phi$), is
\begin{equation}
  \label{eq:57}
  I_{\nu_0}(\vec{r},\theta,\phi) = \frac{L_{\nu_0}}{4\pi (r^2 +
    d^2)} sin\theta_*\delta(\theta - \theta_*)\delta(\phi-\pi)
\end{equation}
where $sin\theta_* = d/\sqrt{r^2 + d^2}$, $d$ is the binary separation,
and $\delta(x)$ is the Dirac-$\delta$ function.

\begin{figure}
  \centering
\includegraphics[width=0.49\textwidth,
  height=0.2\textheight]{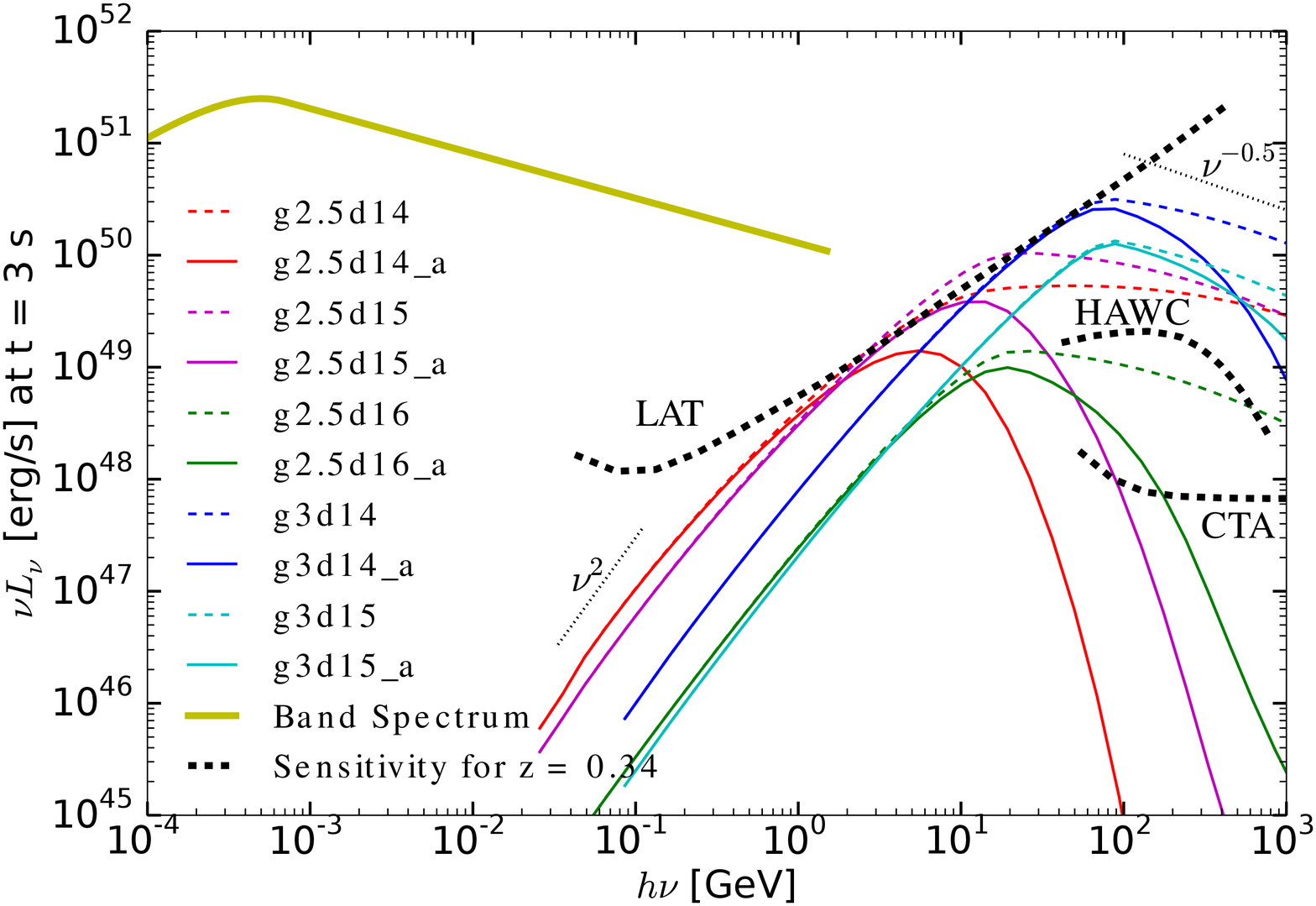}
  \includegraphics[width=0.49\textwidth,
  height=0.2\textheight]{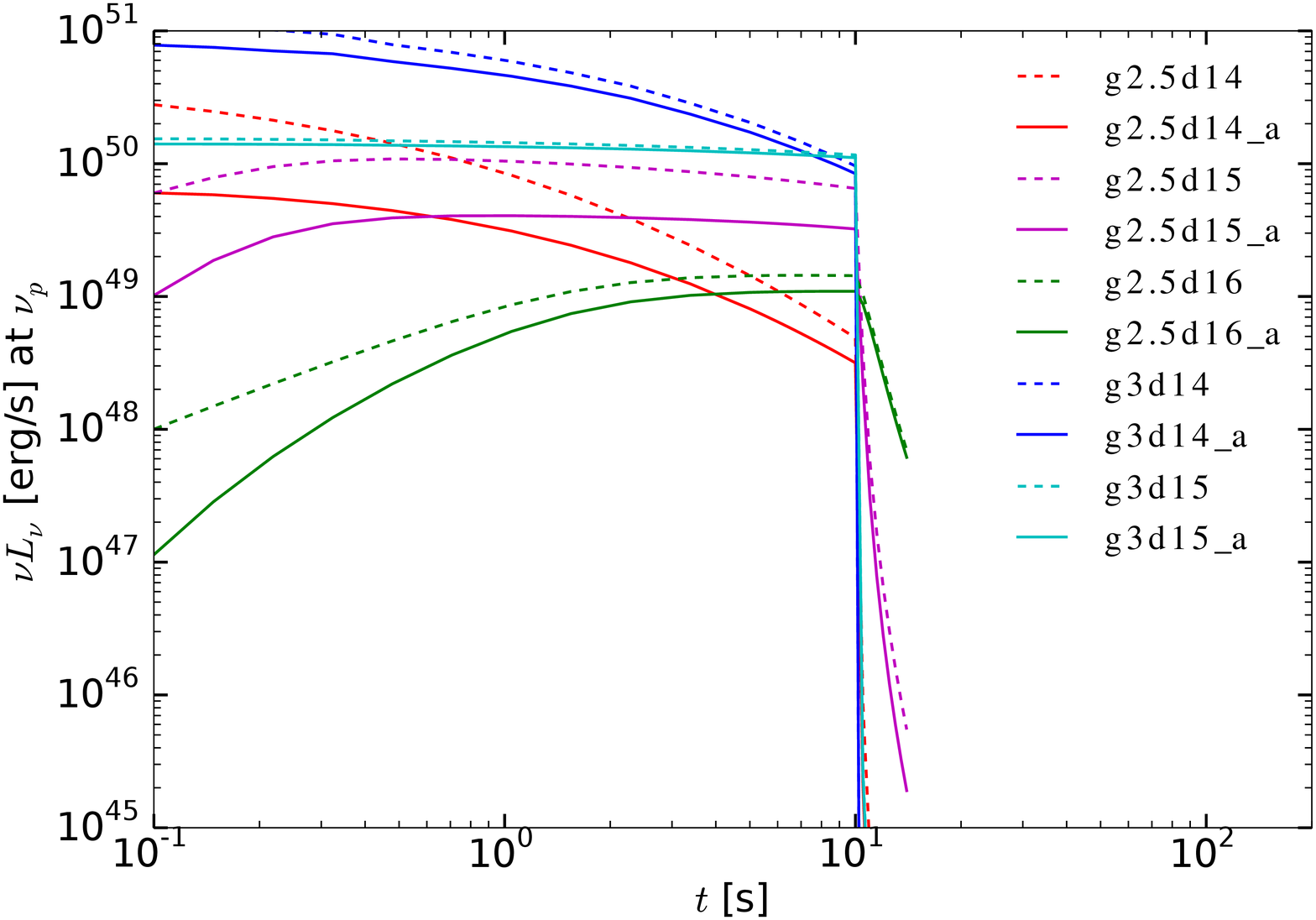}
  \caption{ \small Spectra at
    $t_{obs} = 3\mbox{ }s$ (upper panel) and lightcurves at $\nu_p$ in
    the {\it Binary Case}, for different binary separations 
    $d$ and bulk Lorentz factors $\Gamma$
    (e.g. ``d14'' means $d = 10^{14}\ cm$ and ``g2.5'' means
    $\Gamma=10^{2.5}=300$).  ($\nu_p$'s are the peaks of $\nu L_\nu$
    spectra with absorption, solid lines, in the upper panel.)  We
    show the differences between  
    calculations with/without 
    pair-production absorption by using dashed/solid lines (``\texttt{\_a}''
    means absorption is included). Absorption is weaker for a
    larger bulk Lorentz factor or a 
    larger binary separation. The scaled parameters $E_{j,54}$,
    $T_{j,1}$, $\gamma_{m, 2}$, $T_{4.7}$, $L_{b,39}$ are all set
    to 1 and the electron powerlaw index $p = 2.2$. The thick {\bf
      \textcolor{GreenYellow}{yellow}}
    line is the ``classical'' Band spectrum
    \citep{1993ApJ...413..281B} with $E_{peak} = 500\ keV, 
    \alpha = -1, \beta = -2.4, L_{tot} = 10^{52}\ erg/s$. The thick
    dashed black lines are differential sensitivity of different
    telescopes. The EIC spectra show 
    $\nu L_\nu\propto \nu^2$ below $\nu_p$ and $\propto
    \nu^{\sim-0.5}$ above $\nu_p$, 
    plus additional high energy softening caused by
    pair production.}
  \label{fig:case1_lc_spectr}
\end{figure}

\begin{figure}
  \centering
  \includegraphics[width=0.49\textwidth,
  height=0.2\textheight]{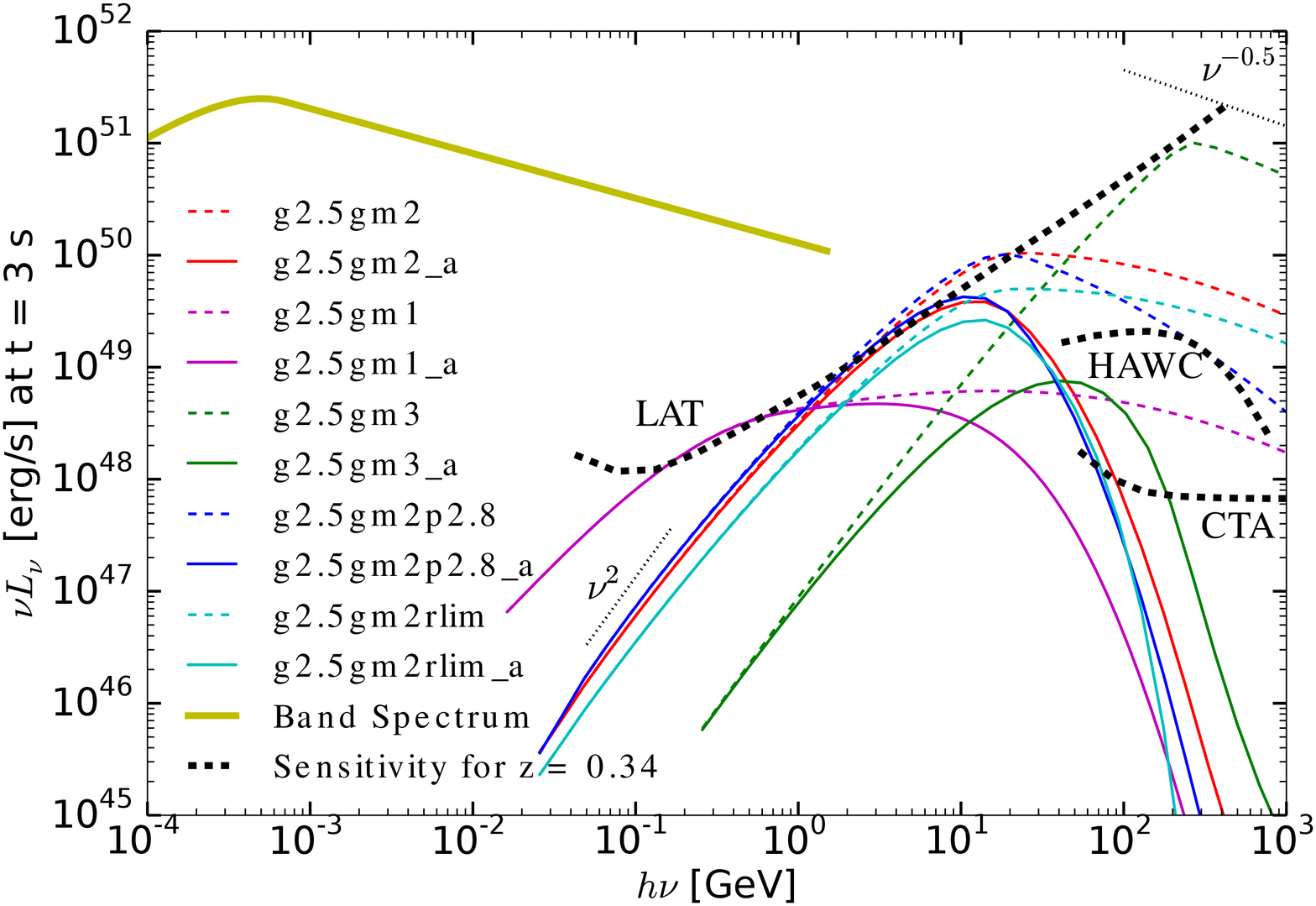}
  \includegraphics[width=0.49\textwidth,
  height=0.2\textheight]{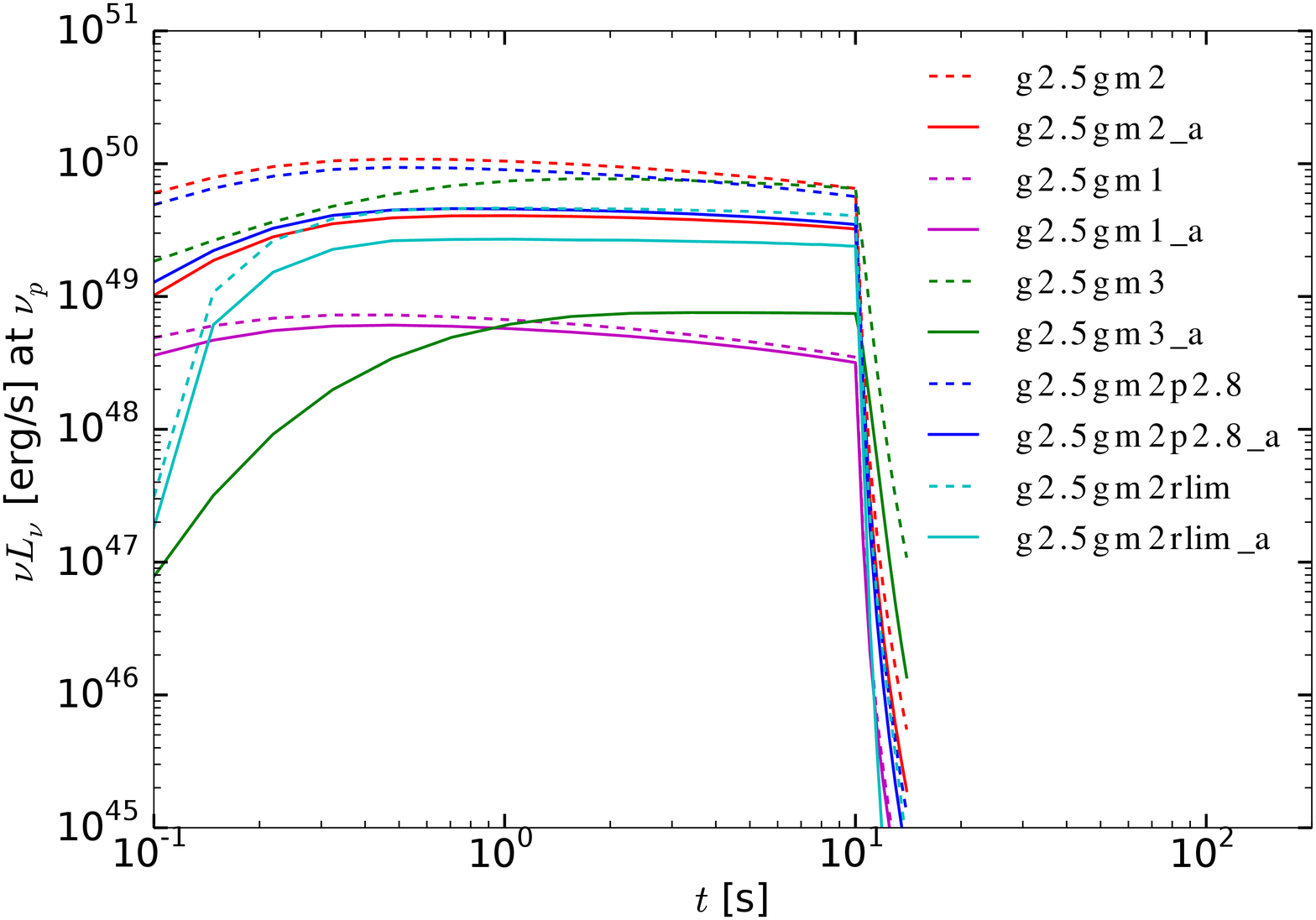}
  \caption{\small Spectra and lightcurves for different jet
    parameters in the {\it Binary Case}. In the fiducial case
    (\textcolor{red}{red}),  $E_{j,54}$, 
    $T_{j,1}$, $\Gamma_{2.5}$, $\gamma_{m, 2}$, $d_{15}$, $T_{4.7}$,
    $L_{b,39}$ are all set 
    to 1 and the electron powerlaw index $p = 2.2$.
    \textcolor{magenta}{Magenta} lines are for $\gamma_m=10$, so the
    scattered photons 
    have lower energies and the un-absorbed spectrum is slightly rising
    from 0.3 to $\sim10$ GeV, because the
    scattering is in the Thomson regime ($\nu L_\nu\propto \nu^{(3-p)/2}$).
    \textcolor{ForestGreen}{Green} lines are for $\gamma_m = 10^3$, so
    the scattered 
    photons reach $\sim 300\ GeV$, but absorption is strong.
    \textcolor{blue}{Blue} lines correspond to $p=2.8$, and the only 
    difference from the \textcolor{red}{red} lines is that the high energy
    spectral slope is steeper, approaching $\nu^{1-p}$ faster because the
    $ln(a_\theta)$ term (Eq. \ref{eq:40}) plays less of a role. For
    \textcolor{cyan}{cyan} lines, we assume electrons are hot only in a
    limited radius range $r\in(d/2,2d)$. The
    small difference between the \textcolor{red}{red} and 
    \textcolor{cyan}{cyan} lines means that, by using a constant
    $\gamma_m$ at all radius, we are making 
    an assumption that electrons are hot near the radius where most
    scatterings happen (see $r_{eff}$ in Table \ref{tab:allcases}). The  {\bf
      \textcolor{GreenYellow}{yellow}} line in the upper panel
    is the ``classical'' Band spectrum (see
    Fig.\ref{fig:case1_lc_spectr} for details) shown for comparison.} 
  \label{fig:case1_rlimited}
\end{figure}

{\it (II) Wind Case}: Assuming a steady wind with mass loss rate $\dot{M}$
and speed $v$, we get the number density of circum-stellar electrons
at a distance $r$ from the center
\begin{equation}
  \label{eq:13}
  n_e(r) = \frac{\dot{M}}{4\pi r^2 v}
\end{equation}
Then the emissivity from wind
scattering is 
\begin{equation}
  \label{eq:14}
  j_{\nu_0}(r) = \frac{n_e(r)\sigma_T}{4\pi r^2}\cdot L_{\nu_0}
\end{equation}
which is assumed to be isotropic (no dependence on $\theta$ or $\phi$) from Thomson
scattering. The system is spherical symmetric, so $j_{\nu_0}$ 
doesn't depend on $\theta_j$ or $\phi_j$, and hence neither does the
intensity $I_{\nu_0}$. We integrate along the 
($\theta,\phi$) direction and get the EPF intensity at position
$\vec{r}$
\begin{equation}
  \label{eq:15}
\begin{split}
&  I_{\nu_0}(r,\theta, \phi) = \int_0^\infty j_\nu(r(s))ds\\
 &=8.0\times10^9\mbox{ } \frac{\dot{M_5}}{v_8}\frac{\pi - \theta + sin\theta
   cos\theta}{sin^3\theta} \frac{L_{\nu_0}}{r^3}
\end{split}
\end{equation}
which is independent of $\phi$, due to symmetry. Note that the
function $f(\theta) = (\pi -
\theta + sin\theta cos\theta)/sin^3\theta \simeq
2(\pi-\theta)^2/3 \rightarrow 0$ when $\theta\rightarrow \pi^{-}$ and
$f(\theta)\simeq \theta^{-3} \rightarrow \infty$
(diverges\footnote{The divergence is caused by the assumption of the
star being a point source. If we denote the progenitor star's radius as
$R_*$, at an angle $\theta<arcsin(R_*/r)$, photons come directly from
the star. However, 
as long as the star is not a giant ($R_*<10^{13}\mbox{ }cm$), the
photons directly from the star are moving nearly parallel to the jet
and are hence negligible.}) when
$\theta\rightarrow 0^{+}$. However, photons with $\theta<1/\Gamma$ are
moving nearly parallel to the jet and hence don't contribute
significantly to the spectra.

{\it (III) Star Cluster Case}: All the stars in the cluster
create a more or less uniform and isotropic EPF, so the intensity can
be estimated as
\begin{equation}
  \label{eq:3}
  I_{\nu_0} \simeq \frac{n_* L_{\nu_0}R_c}{4\pi}
\end{equation}
where $n_*$ is the number density of stars and $R_c$ is the radius of
the cluster.

\begin{figure}
  \centering
  \includegraphics[width=0.49\textwidth,
  height=0.2\textheight]{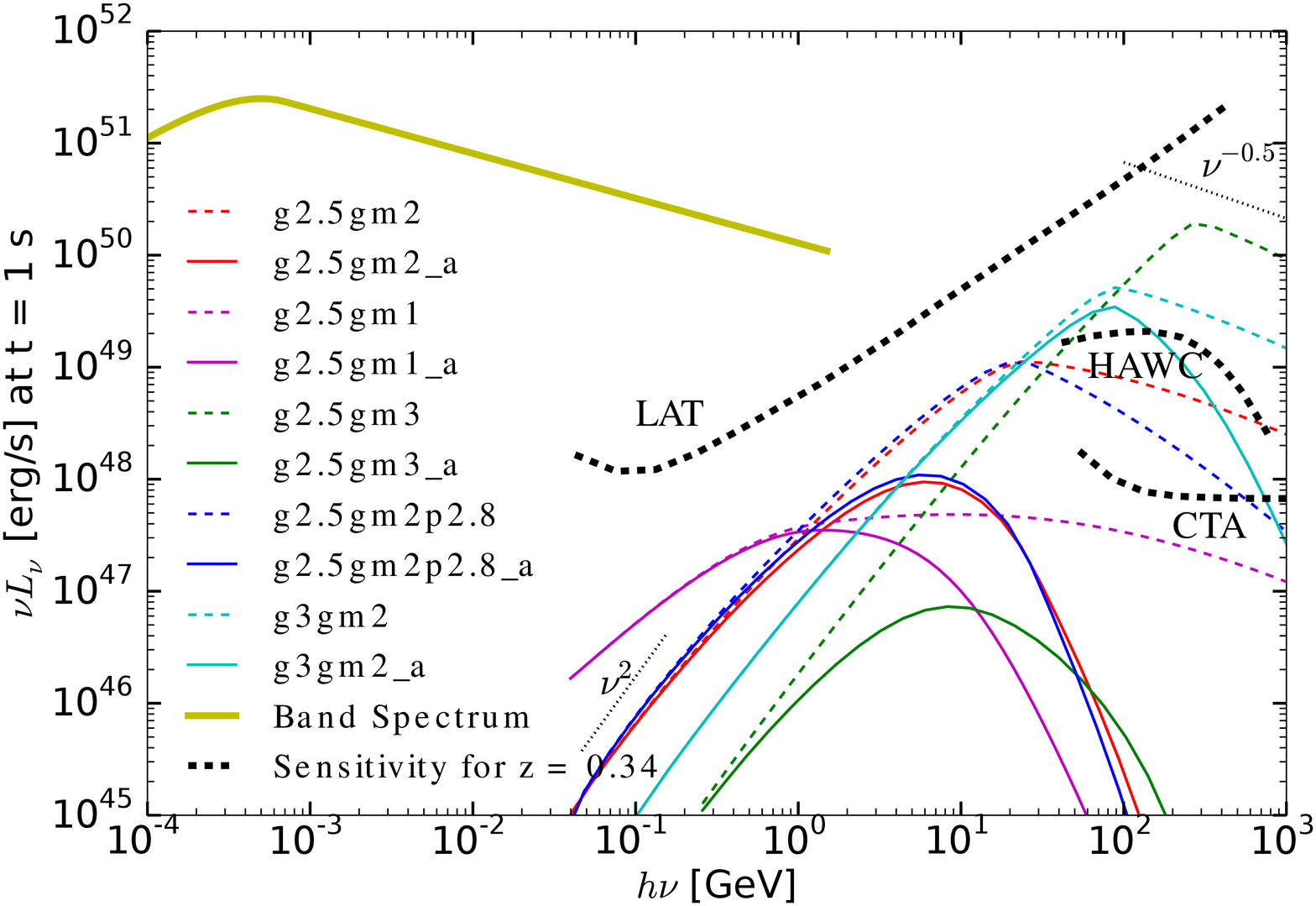}
  \includegraphics[width=0.49\textwidth,
  height=0.2\textheight]{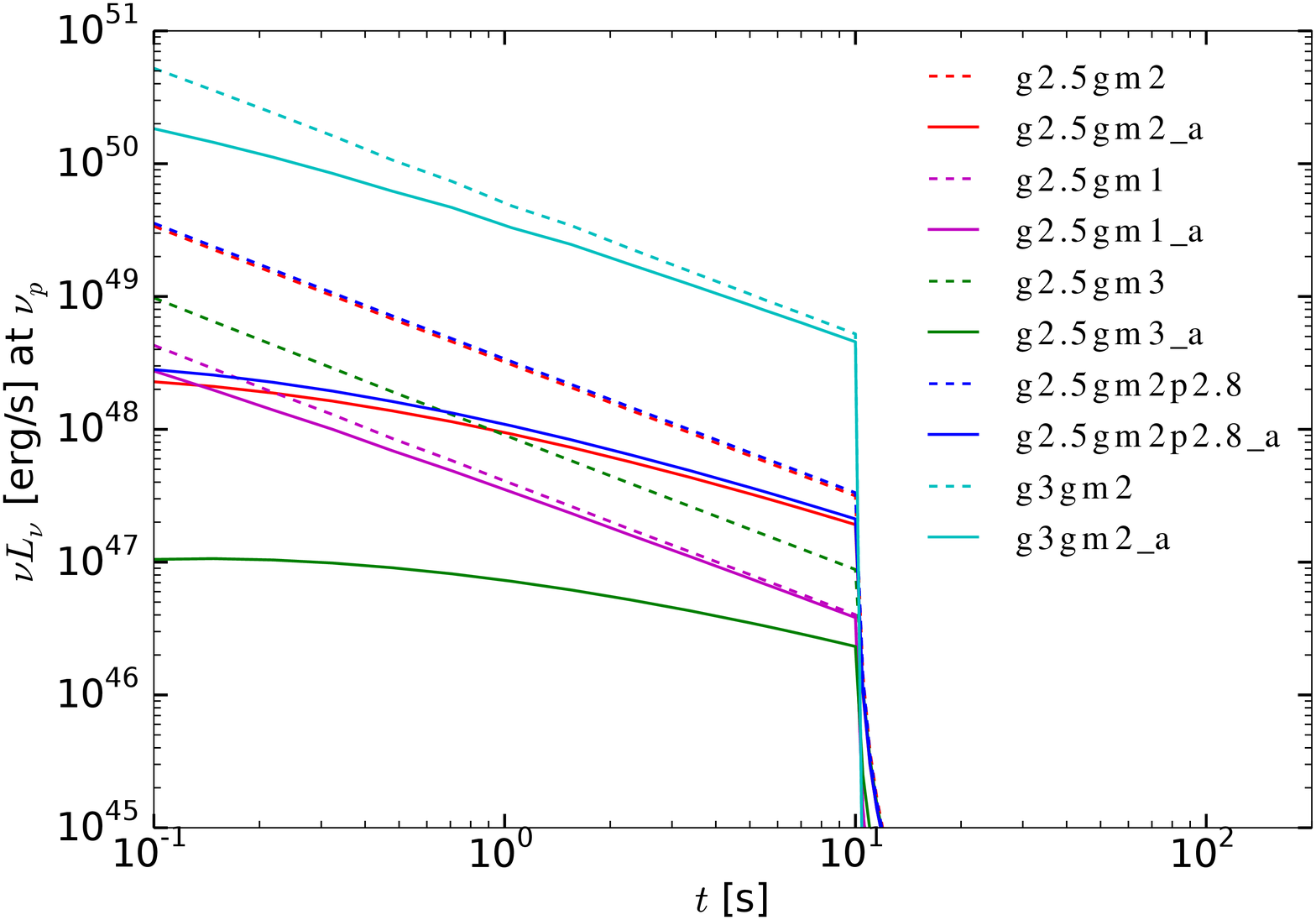}
  \caption{\small Spectra and lightcurves for different jet
    parameters in the {\it Wind Case}. In the fiducial case
    (\textcolor{red}{red}), $E_{j,54}$, 
    $T_{j,1}$, $\Gamma_{2.5}$, $\gamma_{m, 2}$, $T_{4.7}$,
    $L_{b,39}$, $\dot{M}_{-5}$, $v_{8}$ are all set 
    to 1 and the electron powerlaw index $p = 2.2$.
    \textcolor{magenta}{Magenta} lines correspond to $\gamma_m=10$, and
    the scattered photons have lower energies.
    \textcolor{ForestGreen}{Green} lines are for $\gamma_m = 10^3$, so
    the scattered photons reach $\sim 300\ GeV$ but the absorption is
    strong. \textcolor{blue}{Blue} lines correspond to $p=2.8$, and the only 
    difference from the \textcolor{red}{red} lines is that the high energy
    spectral slope is steeper, approaching $\nu^{1-p}$ faster because the
    $ln(a_\theta)$ term (Eq. \ref{eq:40}) plays less of a role.
    \textcolor{cyan}{Cyan} lines are for $\Gamma = 1000$. The  {\bf
      \textcolor{GreenYellow}{yellow}} line in the upper panel
    is the ``classical'' Band spectrum (see
    Fig.\ref{fig:case1_lc_spectr} for details) shown for comparison.}
  \label{fig:allcases}
\end{figure}

\begin{figure}
  \centering
  \includegraphics[width=0.49\textwidth,
  height=0.2\textheight]{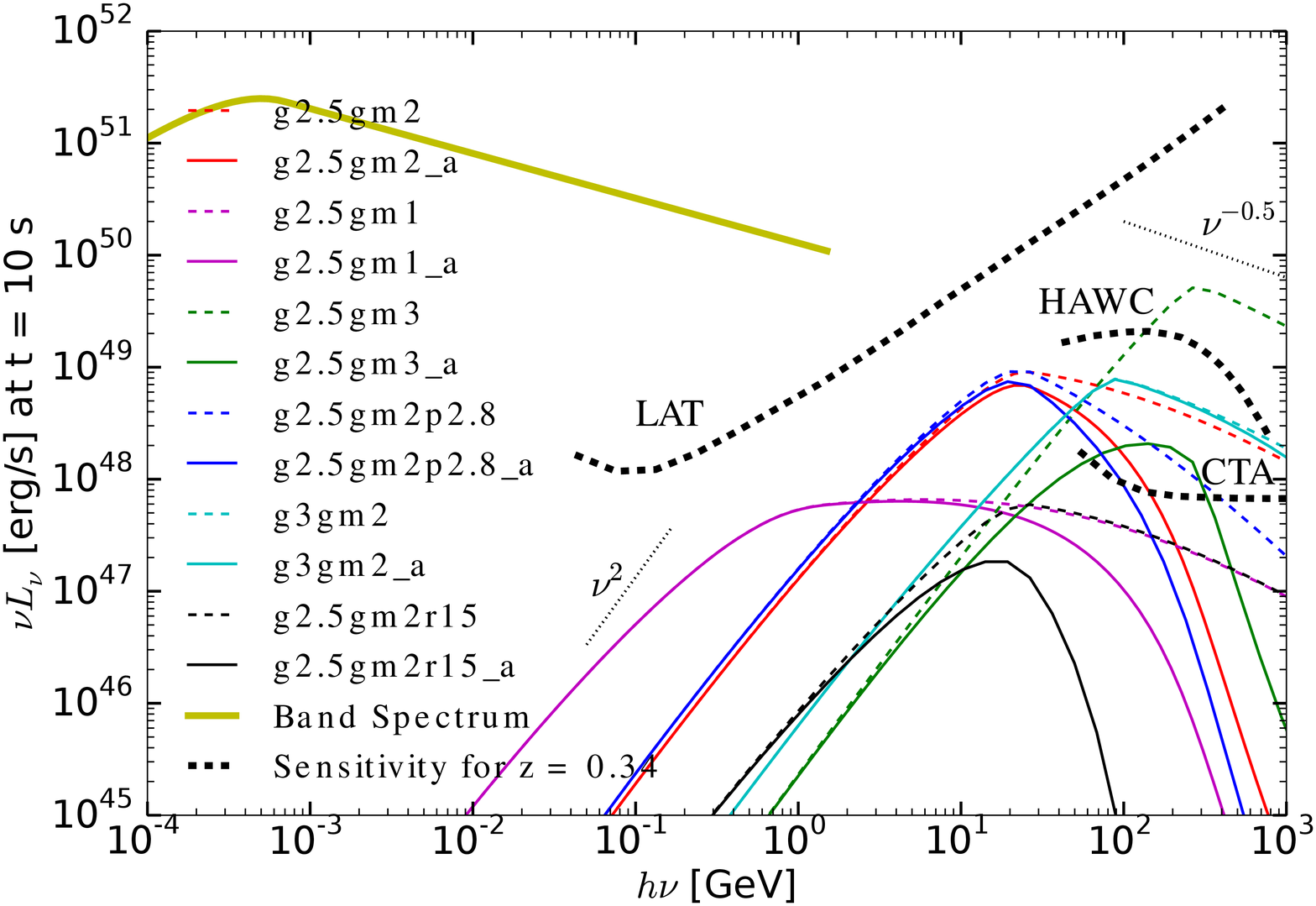}
  \includegraphics[width=0.49\textwidth,
  height=0.2\textheight]{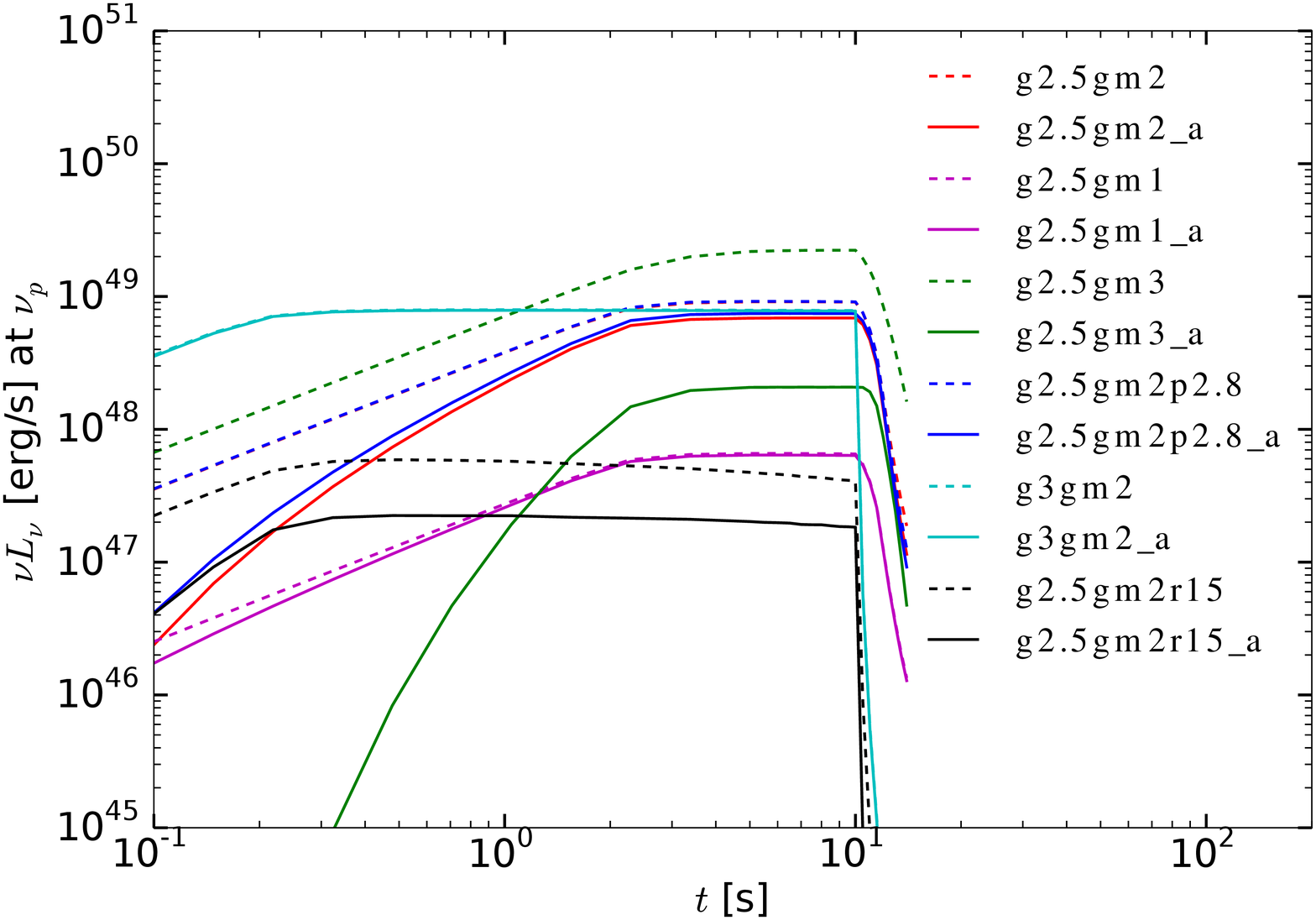}
  \caption{\small Spectra and lightcurves for different jet
    parameters in the {\it Star Cluster Case}. In the fiducial case
    (\textcolor{red}{red}), $E_{j,54}$, 
    $T_{j,1}$, $\Gamma_{2.5}$, $\gamma_{m, 2}$, $T_{4.7}$,
    $L_{b,39}$, $r_{max,16}$, $n_{*,4}$, $R_{pc}$ are all set 
    to 1 and the electron powerlaw index $p = 2.2$.
    \textcolor{magenta}{Magenta} lines correspond to $\gamma_m=10$, so
    the scattered photons 
    have lower energies and the un-absorbed spectrum is slightly rising
    from 0.3 to $\sim10$ GeV, because the
    scattering is in the Thomson regime.
    \textcolor{ForestGreen}{Green} lines are for $\gamma_m = 10^3$, so
    the scattered 
    photons reach $\sim 300\ GeV$, but the absorption is strong.
    \textcolor{blue}{Blue} lines correspond to $p=2.8$, and the only 
    difference from the \textcolor{red}{red} lines is that the high energy
    spectral slope is steeper, approaching $\nu^{1-p}$ faster because the
    $ln(a_\theta)$ term (Eq. \ref{eq:40}) plays less of a role.
    \textcolor{cyan}{Cyan} lines are for $\Gamma = 1000$. For black
    lines, the maximum radius that  
    electrons stay hot is $r_{max} = 10^{15}\ cm$. The  {\bf
      \textcolor{GreenYellow}{yellow}} line in the upper panel
    is the ``classical'' Band spectrum (see
    Fig.\ref{fig:case1_lc_spectr} for details) shown for comparison.}
  \label{fig:case3}
\end{figure}

In Fig.(\ref{fig:case1_lc_spectr})-(\ref{fig:case3}), we present the
lightcurves and spectra for the three cases, assuming electrons are
accelerated by internal dissipation. Then we discuss the prospects of
detecting the predicted high energy photons.
We show the representative spectra
  (upper panels)
at observer's time $t_{obs} = 3\mbox{ }s$ for {\it Case (I)},
$t_{obs}=1\ s$ for {\it Case (II)}, and $t_{obs} = 10\mbox{ }s$ for
{\it Case (III)}. Since the overall flux levels of the spectra change with time
(but the
shapes change very little), the different $t_{obs}$'s we have chosen
are when the spectra have their average amplitudes. By integrating the
representative spectra over the duration of EIC emission $\sim T_j = 
10\ s$, we obtain the total fluence approximately. 
The jet parameters $E_{j,54}=1$,   $T_{j,1}=1$, 
$\Gamma_{2.5}=1$, $\gamma_{m,2}=1$ and $p=2.2$ are our fiducial values
in all three cases. To explore the effect of varying jet parameters on
EIC emission, we also show the
results for a number of different parameters (one at a time): $\Gamma
= 1000$ $(\Gamma_3 = 1)$, $\gamma_m = 10$ $ (\gamma_{m,1} = 1) $, $
\gamma_m =  1000$ $(\gamma_{m,3} = 1)$ and $p=2.8$. For each set of
parameters, the solid (dashed) line is when pair-production absorption
with GRB prompt emission is (not) considered. Then we measure the peak
frequency $\nu_p$ of the solid line (with absorption) for each set
of parameters. The lightcurves (shown in lower panels) are calculated at
$\nu_p$. In this way, the solid-line lightcurves are close to
bolometric and we can see how strong the absorption is at $\nu_p$.

To demonstrate the detectability of the IC emission, (1) along
  with the EIC spectra,  we also plot the ``classical'' prompt
  emission, i.e. Band spectrum \citep{1993ApJ...413..281B}, with
  $E_{peak} = 500\ keV, \alpha = -1, \beta = -2.4, L_{tot} = 10^{52}\
  erg/s$. It's still an open question whether the Band spectrum
  extends to a few GeV without a cutoff. In some cases, the
  EIC emission is only observable if Band spectrum has a GeV
  cutoff. (2) We show in
  Fig.(\ref{fig:case1_lc_spectr})-(\ref{fig:case3}) the sensitivity of
  {\it Fermi} Large Area Telescope 
  \citep[LAT,][]{2009ApJ...697.1071A}, High Altitude Water Cherenkov
  \citep[HAWC,][]{2012APh....35..641A} observatory, Cherenkov Telescope
  Array \citep[CTA,][]{2013APh....43..252I} with integration time $10\
  s$ for a GRB at redshift z = 0.34, which is the redshift of the
  nearest LAT detected GRB \citep[130427a,][]{2014Sci...343...42A}.

  The sensitivity of {\it Fermi} LAT is
  calculated using $\nu F_\nu|_{min} = h\nu 
  N_{ph}/(A_{eff}(\nu) *t_{intg})$, where $N_{ph}=1$ is the minimum
  number of photons, $t_{intg}=10\ s$ is the integration time, and
  $A_{eff}(\nu)$ is the detector's effective area
  \citep[for ``source'' class photons at 
  normal incidence,][]{2009ApJ...697.1071A}. Considering
  background and non-normal incidence angles, the
  true sensitivity of {\it Fermi} LAT is worse than shown
  here. We can see that {\it Fermi} LAT doesn't have enough
  collecting area to effectively constrain the EIC emission
  parameters for GRB with $z\gtrsim 0.34$. However, HAWC and CTA
  will be sensitive enough.

  \citet{2015ApJ...800...78A} calculate the HAWC (-300) sensitivity with
  integration time $t_{intg}=21.5\ s$, and we scale their results for
  the main data acquisition system by $t_{intg}^{0.7}$ and obtain the
  sensitivity for $10\ s$ intetration time. As for the CTA sensitivity, we
  follow the procedure of \citet{2013APh....43..171B} and calculate
  the minimum flux level of CTA South that satisfies the following three
  conditions: significance level $\sigma\geq 5$, there are at least 10
  signal events, and the signal is at least 5\% of the remaining
  background in the direction of the source.

A summary of our numerical results is as follows:

(1) EIC emission is concentrated in $1$ - $100\mbox{ }GeV$
band. The EIC energy is typically $10^{48}\sim10^{51}\mbox{
}erg$ (hence luminosity $10^{47}-10^{50}\mbox{
}erg/s$ ) and the photon fluence is $10^{-7}\sim10^{-4}\mbox{ 
}cm^{-2}$ for a typical redshift\footnote{The luminosity distance $D_{L,z=1}
  =2.0\times10^{28}cm$ from $D_L = c(1+z)/H_0\int_0^z
dx/\sqrt{\Omega_m(1+x)^3 + \Omega_\Lambda}$, and $\Omega_m = 0.27$,
$\Omega_\Lambda = 0.73$, $H_0 = 71 \mbox{ }km\mbox{ }s^{-1}Mpc^{-1}$.}
$z = 1$.

(2) The numerical spectra follow the scaling of the averaged
differential cross section (Eq.\ref{eq:38}) quite well. Without
considering pair production, we get $\nu L_\nu\propto \nu^2$ below
$\nu_p$ and $\propto \nu^{\sim-0.5}$ above $\nu_p$. In the cases where
$\gamma_m = 10$, a short segment of $\nu L_\nu\propto \nu^{(3-p)/2}$
can be seen, because the scattering is in the Thomson regime. At
higher energy, the spectra become increasingly steeper and the
expected $\nu^{1-p}$ slope only shows at 
$\sim TeV$. However, pair production usually makes the high energy tail
much softer.

(3) The observational challenge is that the EIC emission from the
internal dissipation arrives within $\sim 10\mbox{ }s$ of the GRB trigger,
which requires that the GRB occurs in the large field of view of a
telescope, e.g. {\it Fermi} LAT, HAWC and
wide-field-mode CTA. However, in the {\it Star Cluster Case},
although not shown in the figures,
EIC photons from the reverse shock will arrive at deceleration time
$t_{dec} = r_{dec}/(2c\Gamma_{{dec}}^2)\sim 200\mbox{ }s$ and those
from the forward shock will
arrive at the time when the scattering changes from KN to Thomson regime
$t_{KN} = r_{KN}/[2c\Gamma_{sh}(r_{KN})^2]\sim2\times10^4\mbox{
}s$. This gives enough time for some
telescopes, e.g. CTA \citep{2013APh....43..252I},
MAGIC \citep{2007ApJ...667..358A},
H.E.S.S. \citep{2009A&A...495..505A},VERITAS
\citep{2011ApJ...743...62A}, to carry out follow-up observations.

(4) In the {\it Binary Case}, the strongest EIC emission is expected when
$d \sim 10^{14}-10^{16}\mbox{ }cm$. Therefore, observations can
probe the existence and properties of a possible luminous companion in
this distance interval, as long as electrons are hot at radius 
$r\sim d$. Many GRBs have been detected in the $0.1-300\ GeV$
band \citep{2013ApJS..209...11A}, but the radiation mechanism of the
GeV photons during the early prompt emission phase is still uncertain
\citep{2015PhR...561....1K}, so the 
observed data points can be used as upper limits unless different
components can be separated and subtracted. Among all the low 
redshift GRBs with LAT GeV data, GRB 130427a 
\citep[z = 0.34,][]{2014Sci...343...42A} provides the most stringent
constraints. However, for the most favorable jet parameters, the
quantum limit of LAT is still slightly above the predicted
fluxes. Therefore, the constraints from LAT are weak. For example, assuming
jet parameters and binary separations as in \textcolor{red}{red},
\textcolor{magenta}{magenta} and \textcolor{blue}{blue}
lines in Fig. \ref{fig:case1_lc_spectr},  we can rule out a 
companion star with luminosity $>5\times 10^{39} \ erg\ s^{-1}$.
On the other hand, HAWC observatory should have been able to put
strong constraints on the external photon field around the progenitor
of GRB 130427a. Unfortunately, the burst occurred under unfavorable
observational conditions \citep[much smaller effective area,
][]{2015ApJ...800...78A}. For example, non-detections under normal
observational conditions would rule out a companion star of luminosity
$>3\times 10^{38} \ erg\ s^{-1}$ at $d\in (10^{14}, 10^{15} cm)$, if
$\Gamma \geq \sim 500$ and $\gamma_{m} \simeq 100$.

We note that, in a possible binary Population III (Pop III)
star system, the luminosity of the companion star could be as high as
$10^{40}\mbox{ }erg/s$ \citep{2002ApJ...567..532H}. A Pop III GRB could
also have a higher jet kinetic energy $E_{j,54} = 10$
\citep{2011ApJ...726..107S}. If a GRB happens in a Pop III binary
system\footnote{The multiplicity and binary separation
  distribution of Pop III stars are largely unknown. By
  simulating the growth and evolution of Pop III stellar
  systems in a sample (N = 10) of minihalos,
  \citet{2013MNRAS.433.1094S} find a binary 
  fraction of $\sim35\%$ (at birth) and that the binary separation
  (at birth) peaks at $10^{2}\mbox{ }AU$. The subsequent
  evolution of these binary systems is still unexplored.} where the
separation between stars is $d = 10^{15}\mbox{ }cm$, for our
fiducial parameters, we expect  
$N_{EIC}^{iso}\simeq 2\times10^{53}$ at $\nu_p/(1+z)\sim 8/(1+z)\mbox{ }GeV$,
which means a photon fluence of $\sim10^{-7}\mbox{ }cm^{-2}$ at
redshift $z = 15$ and that could be observed by future large space
telescopes. 

(5) In the {\it Wind Case}, EIC emission is usually weak
($E_{EIC}\sim10^{49}\mbox{ }erg$), and observations by LAT are not very
constraining. Since $E_{EIC}\propto L_b 
\dot{M}$ (Eq.\ref{eq:44}), this case is worth considering because of
two possible enhancement channels. First, there might be a persistent
(months to years) and strong ($10^{-3}\sim 1\mbox{ }M_{\odot}/yr$)
pre-burst mass loss, as seen in some Type Ibc and all Type IIn supernovae
\citep[SNe, e.g.][]{2007ApJ...657L.105F, 2009ARA&A..47...63S,
  2012ApJ...744...10K}. We also note that the strong mass loss might
not be isotropic\footnote{For example, due to the fast rotation of the
  progenitor, most of the gas could be ejected near the equatorial plane.}
and hence could leave no footprints on the afterglow dynamics. Second,
an unusually high stellar luminosity might be 
produced by, e.g. a SN slightly earlier than the GRB, or a hot
cocoon\footnote{See 
  \citet{2014MNRAS.445..528K} for the EIC scattering of the hot cocoon
  radiation by the GRB jet.} surrounding the jet.

(6) In the {\it Star Cluster Case}, the EIC emission from internal
dissipation is weaker than
the {\it Binary Case} by a factor of $\sim10$. We emphasize the
importance of EIC radiation from the external shock (ES), because we
have a better understanding of the ES than of the internal dissipation
in that the standard (low energy) GRB afterglows are most likely due to 
synchrotron radiation from the ES-accelerated electrons
\citep[e.g.][]{2004RvMP...76.1143P}. However, the uncertainties of the
EIC emission from ES come from the jet dynamical evolution, which
depends on circum-burst number density and jet power. Also, GeV
emission from other mechanisms needs to be separated,
e.g. synchrotron and synchrotron self-Compton emission from forward
shock \citep[see][for a review]{2015PhR...561....1K}. We predict
photon fluence from $10^{-6}\mbox{ }cm^{-2}$ (forward shock, $\sim
30\mbox{ }GeV$)  to $10^{-5}\mbox{ }cm^{-2}$ (reverse shock, $\sim 10
\mbox{ }GeV$) at redshift $z=1$, using our fiducial parameters. Future
large telescopes may be able to constrain the properties of the cluster.

\section{Discussion}\label{section:discussion}
In this section, we discuss some potential issues of the EIC emission
proposed in this paper.

(1) We note that high energy $\gamma$-rays can't propagate
through cosmological distances, due to pair
production with extragalactic background light (consisting mostly of
cosmic infrared, optical and UV backgrounds). Photons of energy
$\geq100 \mbox{ }GeV$ can only be observed from nearby ($z\lesssim1$)
GRBs 
\citep{2012MNRAS.422.3189G}. Therefore, constraints on the
  external photon field will be
most effective for low redshift ($z\lesssim 1$) GRBs. We also note
that GRBs with low bulk Lorentz factor $\Gamma\lesssim300$ will mostly produce
$\lesssim 10 \  GeV$ photons, which could be observed at high redshift.

(2) We assume the jet kinetic power to be steady for a duration
$T_j$. In reality, GRB prompt lightcurves show multiple
emission pulses. Some pulses are smooth but others vary on
millisecond timescale, so the jet kinetic power could be episodic and
the internal dissipation could happen locally. The
EIC emission is also modulated by the radial distribution of external
photon field. If
most EIC scatterings happen at radii $\sim r$, an infinitely thin shell will
produce a pulse of width $\sim 0.2s\ r_{15}/\Gamma_{2.5}^2$. Due to limited
number of EIC photons from a source located at cosmological distances,
telescopes are not sensitive to the variability 
down to milliseconds. Instead, the observed EIC lightcurve should more
or less track the major pulses of the prompt emission, with smoothing
on timescales of order $\sim 0.2s\ r_{15}/\Gamma_{2.5}^2$.

(3) In the {\it Binary Case}, the EIC luminosity
is only significant when the binary separation is large ($d\geq
10^{14}\mbox{ }cm$). Orbital separation distribution 
derived from spectroscopic 
and  direct imaging studies are inevitably biased by selection effects
and measurement limitations. By measuring
radial velocities, \citet{2012Sci...337..444S} derived (after
completeness corrections) the distribution of orbital 
periods $f(log P)\propto (log\mbox{ }P)^{-0.55}$ for $P=1.4-3200\mbox{
}d$ by analyze the O star population of six nearby Galactic open
clusters. Also by measuring radial velocities,
\citet{2014arXiv1406.6655K} estimated $f(log\mbox{ }P)\propto (log\mbox{
}P)^{-0.22}$ for $P=1.4-2000\mbox{ }d$ in the Cygnus OB2
Association. Despite the discrepancy (and also uncertainties),
both of them pointed out that about $10\%$
of the binary systems have periods $P> 10^3\mbox{ }d$,
which corresponds to $d\simeq10^{14}\mbox{ }cm$. An even greater
uncertainty comes from the unknown nature of GRB progenitors, which
may not be O stars \citep{1993ApJ...405..273W,
1999ApJ...524..262M}. Detection of the EIC emission proposed in this
paper will shed light on the nature of GRB progenitors.

(4) In the {\it Star Cluster Case}, the EIC luminosity depends on the
number density of massive stars at the time when the GRB occurs. There
are large uncertainties in the first few Myr of the star clusters'
evolution \citep[see e.g.][]{2010ARA&A..48..431P}, especially in the
so-called Gas Expulsion  Stage (GES) during which the residual gas in
the cluster is blown away by stellar winds or outflows. The loss of
mass during GES will cause the cluster to expand and the shallower
potential well may allow a fraction of the stars (especially the ones
at outer radius) to
escape. Therefore, the overall stellar number density drops with
time. Long GRBs are concentrated on the very brightest regions of
actively star-forming galaxies, where young massive
clusters are particularly abundant
\citep{2006Natur.441..463F}. However, little is known about the 
very nearby ($\sim 1\mbox{ }pc$) environment around the long GRB 
progenitors, and it is unclear whether the stars
escaped\footnote{\citet{2006A&A...454..103H} found GRB 980425
\citep[the nearest GRB at $z=0.008$, e.g.][]{1998Natur.395..670G} to be
$\sim 800 \mbox{ }pc$ away from the nearby massive star-forming
region.} from the clusters. The EIC signal proposed in this 
paper could potentially give us some hint.
For example, an upper limit of $L_{EIC}<10^{49}\mbox{ }erg/s$ will
constrain the number density of massive ($>20$ -- $30M_\odot$) stars
to be $n_*<10^4\mbox{ }pc^{-3}$. 

(5) There are a few potential issues in our numerical procedures
developed in Section \ref{section:theory}. (a) We assume electrons are
accelerated to ultra-relativistic energies and have a simple powerlaw
Lorentz factor (LF) distribution. On one hand, if GRB prompt
emission is produced via sub-photospheric Comtonization, then
$\gamma_m$ (or electrons' thermal LF) might be of order a few or
less. In that case, the EIC emission would be much weaker than the
internal dissipation model considered in this work. On the other
hand, even if electrons are accelerated to a simple powerlaw, due to
synchrotron, synchrotron-self-Compton, or EIC cooling, 
there's a cooling LF $\gamma_c$ above which electrons' distribution turns 
softer. The resulting EIC spectrum at energy higher than
$\Gamma\gamma_c m_ec^2$ will be softer than our prediction, but the
total EIC energy will not change much, as long as $\gamma_c>\gamma_m$.
(b) Electron's LF distribution is actually evolving with time, i.e.
$\gamma_m$, $\gamma_c$ and possibly the powerlaw index $p$ are all
functions of time (or radius). In principle, by convolving the time
dependent LF distribution with 
single particle emission, we can calculate the EIC emission to a
higher degree of accuracy. However, since we are not yet sure where the
internal dissipation accelerates electrons and how long will
electrons stay hot \citep{2015PhR...561....1K}, such a detailed
calculation is left for possible future investigation. Instead,
constant $\gamma_m$ and $p$ are used at all radius. As shown in
Section \ref{lc_spectr} (Fig.\ref{fig:case1_rlimited}),  
by using a constant $\gamma_m$ at all radius, we are making an
assumption that electrons are hot near the radius $r_{eff}$
where most scatterings happen.

(6) Another interesting application is the IC scattering of the Cosmic
Microwave Background (CMB) at high redshift by Pop III GRB
jets. The earliest Pop III stars form in minihalos of mass
$\sim10^{6}M_{\odot}$ at redshift $z\sim20$ - $30$
\citep{2013RPPh...76k2901B}. The CMB photon number density is 
\begin{equation}
  \label{eq:62}
  n_{CMB} \simeq aT^4/(2.7kT)= 3.3\times10^6 [(1+z)/20]^3 cm^{-3}
\end{equation}
where we have used the CMB temperature $T = 2.7(1+z)=54K\mbox{
}(1+z)/20$ (i.e. $\epsilon_0=4.7\times10^{-3}\mbox{
}eV$). The forward shock (FS) is highly relativistic at deceleration
radius, and electrons are accelerated to LF $\gamma_e=1.8\times
10^{5}$ (using  $\epsilon_e = 0.1$, and $\Gamma_{dec} =
100$). Therefore, EIC emission from the FS could be observed at
$0.8GeV\mbox{ }\Gamma_{dec,2}^4$. If we use a jet kinetic energy
$E_{j} = 10^{55}\mbox{ }erg$ (isotropic equivalent) and a 
uniform circum-burst medium density of $n=100\mbox{ }cm^{-3}$, we get
the deceleration radius $r_{dec}\simeq 10^{17}\mbox{ }cm$ (Eq.\ref{eq:9}). 
The total (isotropic) number of EIC photons from FS is
\begin{equation}
  \label{eq:4}
  \begin{split}
      N_{EIC}^{FS} &\simeq (\tau_{FS} n_{CMB} 4\pi r^3/3)|_{r = r_{dec}}\\
  &\simeq 3.0\times 10^{52} [(1+z)/20]^3n_2 r_{dec,17}^4
  \end{split}
\end{equation}
Since $N_{EIC}^{FS}$ strongly depends on $r_{dec}$, a more accurate
calculation of the dynamical evolution is needed.
Using $N_{EIC}^{FS} =3\times10^{52}$, we expect a photon fluence of
$\sim 5\times 10^{-9}\mbox{ }cm^{-2}$ at $\sim 1\mbox{ }GeV$ from
redshift 20.

(7) Lastly, we estimate the possible contribution of the EIC emission
to the extragalactic $\gamma$-ray 
background (EGB) around $100\mbox{ }GeV$, by the FS channel or ID
channel when GRBs have large LF (e.g. $\Gamma\geq500$). We
optimistically estimate the average (isotropic) EIC energy around
$100\mbox{ }GeV$ from one GRB to be $E_{EIC}= 10^{51}\mbox{ 
}erg$. The average (observed) GRB rate in the local ($z<1$) Universe
is estimated by \citet{2010MNRAS.406.1944W} to be $\bar{\rho} \sim 3
\mbox{ }Gpc^{-3}yr^{-1}$.
We integrate over the local Universe to $R_{max} = 4\mbox{ }Gpc$
and estimate the intensity of EGB contributed by the EIC channel to be
$I_{EGB}\simeq E_{EIC} \bar{\rho} R_{max}/4\pi\simeq 2\times10^{-6}
\mbox{ }MeV cm^{-2} s^{-1} sr^{-1}$.  The EGB at $100\mbox{
}GeV$ observed by {\it Fermi} LAT is $\sim 10^{-4} \mbox{ }MeVcm^{-2} s^{-1}sr^{-1}$
\citep{2010ApJ...720..435A}. Therefore, the EIC channel in GRBs could
(at most) contribute a small fraction of EGB. The possible
contributions from other GRB channels are estimated by
\citet{2007ApJ...656..306C}.
\section{Conclusions}\label{section:conclusions}
GRB progenitors may be surrounded by a significant external photon
field (EPF), due to the existence of a massive 
companion star ({\it Case I}), a strong stellar 
wind ({\it Case II}), or a dense star cluster ({\it Case III}). We
calculate the IC scattering of the EPF (i.e. EIC emission) by the hot
electrons in the GRB jet, which could be accelerated by internal
dissipation or external shocks. As long as electrons are accelerated to
ultra-relativistic powerlaw, the results presented in this work is
independent of the details of jet dissipation mechanism and electron
acceleration process.

In each {\it Case}, the EPF-contributing star(s) are assumed to have
bolometric luminosity $10^{39}\mbox{ }erg/s$ and effective
temperature $5\times10^{4}\mbox{ }K$. The GRB jet is assumed to have
total kinetic energy $10^{54}\mbox{ }erg$ (isotropic equivalent),
duration $T_{j} = 10\mbox{ }s$ and bulk Lorentz factor $\Gamma$. We
assume that electrons are accelerated by internal dissipation to a powerlaw
distribution with minimum Lorentz factor
$\gamma_m$ and index $p$. We present in
this work the EIC lightcurves and spectra for each case, for a variety
of EPF and jet parameters ($\Gamma$, $\gamma_m$ and $p$). We take 
into account 
the equal-arrival time surface and possible absorption of high energy
photons caused by pair production with the prompt $\gamma$-rays.

For our fiducial jet parameters $\Gamma
= 300$, $\gamma_m = 100$ and $p = 2.2$, the EIC spectrum peaks at
$\nu_p\sim10\mbox{ }GeV$. The EIC spectrum below $\nu_p$ is  $\nu
L_\nu\propto \nu^2$ and $\propto  \nu^{\sim-0.5}$ above $\nu_p$; pair
production softens the spectrum at high energies (above the threshold
for $e^\pm$ production). In {\it Case (III)}, significant EIC
emission also comes from electrons accelerated by external shocks,
where we use $\epsilon_e = 0.1$ as the fraction of internal energy
shared by hot electrons in the shocked regions.

In {\it Case (I)}, using binary separations $10^{14}-10^{16}$ $cm$,
we get EIC luminosities of $10^{47}-10^{50}$ $erg/s$ (the effect
of pair production on high energy photon luminosity is included in
this and all calculations below), peaking at separation $10^{15}$
$cm$. In {\it Case (II)}, soft photons originally from the progenitor star
are scattered first by the electrons in the stellar wind and 
again by the jet. Using a W-R-star mass loss rate $10^{-5}\mbox{
}M_{\odot}/yr$ and wind velocity $10^{8}\mbox{ }cm/s$, we get an EIC
luminosity of $10^{48}\mbox{ }erg/s$. In {\it Case (III)}, we assume
that stars in the cluster create a uniform EPF. Electrons in the jet are 
accelerated by both internal dissipation (ID) and external shocks. We
include both the reverse shock and forward shock. Using stellar number
density $10^4\mbox{ 
}pc^{-3}$ and cluster radius $1\mbox{ }pc$, we get EIC luminosities of
$10^{49}\mbox{ }erg/s$ from ID, and $10^{49}\mbox{ }erg/s$ from external
reverse shock and
$10^{46}\mbox{ }erg/s$ from external forward shock. The EIC emission
from ID lasts for $t_{obs}\simeq T_{j}\sim10\mbox{ }s$, but that 
from external shocks lasts longer, approximately $200\mbox{ }s$ from
reverse shock and $2\times10^4\mbox{ }s$ from forward shock.

We note that EIC emission from ID relies on the assumption that
electrons are hot at the right radius $r_{eff}$:
$\sim d$ (the binary separation, {\it Case I}), $\sim r_{tr}$
($\sim3\times10^{14}\mbox{ }cm$, the radius
where the jet becomes transparent, {\it Case II}) and
$10^{16}cm \mbox{ } r_{max,16}$ (the maximum radius electrons stay
hot, {\it Case III}). In addition, EIC emission from external shocks
strongly depends on the dynamical evolution of the jet.

Generally, from a GRB at redshift $z=1$, a photon fluence of
$10^{-7}-10^{-4}\mbox{ }cm^{-2}$ at 
$1-100\mbox{ }GeV$ is expected. Future observations (by e.g. {\it Fermi}
LAT, HAWC and CTA) can put constraints on: (1) the existence of such
EPFs as described in this paper and hence on the nature of GRB
progenitors; (2) the radius where the jet internal dissipation process
accelerates electrons.

\section*{Acknowledgments}
The authors thank M. Milosavljevi\'{c}, P. Crumley, R. Santana,
R. Hern\'{a}ndez for helpful discussions. G.F.S. acknowledges support
through his Chaire d'Excellence Universit́\'{e} Sorbonne Paris Cit\'{e} and
the financial support of the UnivEarthS Labex program at Universit\'{e}
Sorbonne Paris Cit\'{e} (ANR-10-LABX-0023 and ANR-11-IDEX-0005-02).

\appendix
\section{}
There are two potential pair-production channels that could 
absorb the high energy ($10-100\mbox{ }GeV$)
the EIC photons. (1) In the lab frame, the EPF ($\sim10\mbox{ }eV$) will
interact with $\sim100\mbox{ }GeV$ photons. We show below that this
channel is usually not important. (2) In the jet 
comoving frame, the GRB prompt emission ($\sim1\mbox{ }keV$) will
interact with $\sim1\mbox{ }GeV$ photons ($\sim 100\  GeV$ in the
lab frame). We consider this channel in detail, following 
\citet{1996ApJ...456..124M, 2010ApJ...721..709G}, and this channel is
included in our numerical calculation (Section \ref{section:theory}).

\subsection{Absorption by EPF Itself}
The only concern here is for photons of energy
$\geq100\mbox{ }GeV$. The optical depth of pair
production is
\begin{equation}
  \label{eq:11}
  \tau_{\gamma\gamma}\simeq n_{\gamma}\sigma l
\end{equation}
where $l$ is the path length of the high energy $\gamma$-ray through
the EPF, $n_{\gamma}$ is the number density of field photons near the
energy of maximum cross section, and $\sigma\simeq 0.1 \sigma_T$ is
the approximate cross section.

In the {\it Binary Case}, most scatterings
happen at radius $r\sim d$, so we have
\begin{equation}
  \label{eq:23}
      \tau_{\gamma\gamma}^{(1)}\simeq \frac{L_b}{4\pi(d^2 +
    d^2)\epsilon_0c}(1 - \frac{\sqrt{2}}{2})\sigma d
= 1.6\times10^{-2}\frac{L_{b,39}}{d_{14}\epsilon_{0,1}}
\end{equation}
In the {\it Wind Case}, most scatterings happen at radius $r\sim r_{tr}$
(see Eq.\ref{eq:5} and \ref{eq:27}), so we have
\begin{equation}
  \label{eq:24}
     \tau_{\gamma\gamma}^{(2)} \simeq \frac{L_b \tau_{wind}(r_{tr})}{4\pi r_{tr}^2
    \epsilon_0 c}\sigma r_{tr}
=
2\times10^{-6}\frac{\Gamma_{2.5}L_{b,39}\dot{M}_{-5}}{E_{j,54}\eta_{KN}
  \epsilon_{0,1} v_8} 
\end{equation}
In the {\it Star Cluster Case}, EIC photons need to penetrate the whole cluster
that has radius $R = 1pc\mbox{ }R_{pc}$, so we have
\begin{equation}
  \label{eq:33}
     \tau_{\gamma\gamma}^{(3)} \simeq \frac{n_* L_b R}{\epsilon_0
       c}\sigma R
= 4.5\times10^{-1}\frac{n_{*,4}L_{b,39}R_{pc}^2}{\epsilon_{0,1}}
\end{equation}
These are upper limits, since some EPF photons are either offset from the
maximum cross section energy or at energy lower than the
threshold. Anyway, the absorption is only important 
in {\it Binary Case} when $d\leq\sim10^{12}\mbox{ }cm$ and in {\it
  Star Cluster Case} when $R\geq\sim1\mbox{ }pc$. Under these
conditions, there will be a 
spectral cut off at $>\sim 100\mbox{ }GeV$. The absorption is
not important in {\it Wind Case}. A more detailed
calculation is given by \citet{2010ApJ...721..709G}.

\subsection{Absorption by GRB Prompt Emission}
In the jet comoving frame, a high energy EIC photon of frequency $\nup$
is surrounded by low energy photons (frequency $\nu_l^\prime$), which
can be described by the so-called Band function \citep{1993ApJ...413..281B}. We
simplify the Band function as a broken powerlaw of indexes $\alpha_1$
and $\alpha_2$, below and above the break 
frequency $\nu_b$. From the prompt
emission bolometric isotropic luminosity $L_\gamma$, and the radius $r$
where the low energy photons are produced, we get the comoving-frame
specific number density
\begin{equation}
  \label{eq:36}
  \begin{split}
         n^\prime_{\nu_l^\prime}(r) = &\frac{L_\gamma}{4\pi r^2 c h\nu_b^2}
  (\frac{1}{\alpha_1+2} - \frac{1}{\alpha_2+2})^{-1} \cdot\\
&
  \begin{cases}
    (\nu_l^\prime/\nu_b^\prime)^{\alpha_1} &\mbox{ if
    }\nu_l^\prime<\nu_b^\prime\\
   (\nu_l^\prime/\nu_b^\prime)^{\alpha_2} &\mbox{ if
    }\nu_l^\prime>\nu_b^\prime\\
  \end{cases}
  \end{split}
\end{equation}
where $\nu_b^\prime = \nu_b/\Gamma$ is the break frequency in the
comoving frame. The pair-production threshold is
\begin{equation}
  \label{eq:37}
  h\nu_{th}^\prime = \frac{2(m_e c^2)^2}{h\nu^\prime (1-cos\theta_l^\prime)}
\end{equation}
where $\theta_l^\prime$ is the angle between high($\nu^\prime$)
and low($\nu_l^\prime$) energy photons' momentum vectors. The 
cross section for pair production is \citep{1996ApJ...456..124M}
\begin{equation}
  \label{eq:49}
  \sigma_{\ggam}(\nu^\prime,\nu_l^\prime,\theta_l^\prime) =
  \frac{3\sigma_T}{16}(1-q^2) \left[ (3-q^4)ln\frac{1+q}{1-q} + 2q(q^2
    - 2) \right]
\end{equation}
where $q = \sqrt{1-\nu_{th}^\prime/\nu_l^\prime}$. Assuming that low
energy photons are moving isotropically in the jet comoving frame, we
get the mean free path
\begin{equation}
  \label{eq:58}
  l_{\ggam}^{-1}(\nu^\prime,r) =
  \frac{1}{2}\int_{-1}^{1}d(cos\theta_l^\prime) (1-cos\theta_l^\prime)
  \int_{\nu_{th}^\prime}^\infty d\nu_l^\prime n_{\nu_l^\prime}^\prime \sigma_{\ggam}
\end{equation}
Then the pair-production optical depth is the comoving dynamical
timescale divided by the mean time interval between two collisions,
i.e. 
\begin{equation}
  \label{eq:59}
  \tau_{\ggam}(\nu^\prime,r) = \frac{r}{\Gamma l_{\ggam}(\nu^\prime,r)}
\end{equation}
Therefore, only a fraction $e^{-\tau_{\ggam}}$ of the EIC photons in
Eq.(\ref{eq:41}) can escape. A simple estimate of the pair-production
optical depth is
\begin{equation}
  \label{eq:60}
      \tau_{\ggam}\simeq 0.1\sigma_T \frac{L_\gamma/4}{4\pi r^2
        ch\nu_{th}^\prime \Gamma^2}
\frac{r}{\Gamma}
\simeq 0.65\mbox{ }\frac{L_{\gamma,52} (h\nu/10GeV)}{r_{15}\Gamma_{2.5}^4}
\end{equation}
where we assume $1/4$ of all the prompt $\gamma$-rays contribute to
the absorption of high energy photons. We can see: (1) $\tau_{\ggam}$
is very sensitive to the bulk Lorentz factor (we have used
$\Gamma=300\Gamma_{2.5}$); (2) most absorption 
happens at small radius ($r_{15}\leq1$) and high frequency
($h\nu\geq10\mbox{ }GeV$).

\label{lastpage}

\end{document}